\documentclass[amsmath,caption,amssymb,superscriptaddress,showpacs, aps,prfluids,longbibliography]{revtex4-1}
\usepackage{graphicx}

\usepackage{color,soul}
\usepackage{mathrsfs}
\usepackage{todonotes}
\usepackage{bm}

\usepackage{csquotes}

\newcommand{\de}{\frac{\partial }{\partial \eta}}
\newcommand{\dev}[1]{\frac{\partial #1}{\partial \eta}}
\newcommand{\dt}[1]{\frac{\partial #1}{\partial t}}
\newcommand{\dz}[1]{\frac{\partial #1}{\partial z}}
\newcommand{\dr}[1]{\frac{\partial #1}{\partial r}}
\newcommand{\dzz}[1]{\frac{\partial^2 #1}{\partial z^2}}
\newcommand{\dC}[1]{\frac{\partial #1}{\partial C^{(0)}}}
\newcommand{\dCz}[1]{\frac{\partial #1}{\partial C^{(1)}}}

\newcommand{\sop}[1]{#1}

\begin{document}



\title{\sop{Active control of dispersion within a channel with flow and pulsating walls}}

\author{Sophie Marbach}
\affiliation{Laboratoire de Physique Statistique, Ecole Normale Sup\'{e}rieure, PSL Research University, 75005 Paris, France}
\affiliation{Courant Institute of Mathematical Sciences, New York University, New York 10012, USA }
\author{Karen Alim}
\affiliation{Department of Physics, Technical University of Munich, Germany}
\affiliation{Max Planck Institute for Dynamics and Self-Organization, 37077 G\"ottingen, Germany }

\date{\today}
\begin{abstract}
Channels are fundamental building blocks from biophysics to soft robotics, often used to transport or separate solutes. As solute particles inevitably transverse between streamlines along the channel by molecular diffusion, the effective diffusion of the solute along the channel is enhanced - an effect known as Taylor dispersion. Here, we investigate how the Taylor dispersion effect can be suppressed or enhanced in different settings. Specifically, we study the impact of flow profile and active \sop{or pulsating} channel walls on Taylor dispersion. We derive closed analytic expressions for the effective dispersion equation in all considered scenarios providing hands-on effective dispersion parameters for a multitude of applications. In particular, we find that active channel walls may lead to three regimes of dispersion: either dispersion decrease by entropic slow down at small P\'{e}clet number, or dispersion increase at large P\'{e}clet number dominated either by shuttle dispersion or by Taylor dispersion. This improves our understanding of solute transport \textit{e.g.}~in biological active systems such as blood flow and opens a number of possibilities to control solute transport in artificial systems such as soft robotics.
\end{abstract} 
\maketitle
\section{Introduction}

Microscale to nanoscale channels are fundamental building blocks for transport and separation of solutes in their stream. Interconnected channels form the vascular networks of animals, plants, fungi and slime molds - an essential transport system for all higher forms of life~\cite{West:1999}. Moreover, channels form lungs, intestines, the nephron of the kidney and the lymph system \cite{Freund:2012, Cartwright:2009}. Channels are the key building block of microscale hydraulic engineering~\cite{stone2004engineering}, with particular importance for porous media applications~\cite{Alim:2017ev} and in the emerging fields of smart materials~\cite{Beebe:2000} and soft robotics~\cite{Shepherd:2011}. The dynamics of solute transport by flow in these channels is essential to transport signals~\cite{shapiro1988dispersion,Alim:2017}, to power chemical reactions~\cite{Zheng:2017jg} and potentially program complex behavior~\cite{jubin2018dramatic}. To increase the space of accessible behaviors one must go beyond the straight static version of a channel and incorporate the boundaries into design: carving for instance cross-sectionally varying channels~\cite{brenner1980dispersion,siwy2006ion,ajdari2006hydrodynamic,chinappi2018charge}, or driving dynamics with active channel walls~\cite{marbach2018transport}.
 
\begin{figure}[htbp]
\center
\includegraphics[width = 0.9\textwidth]{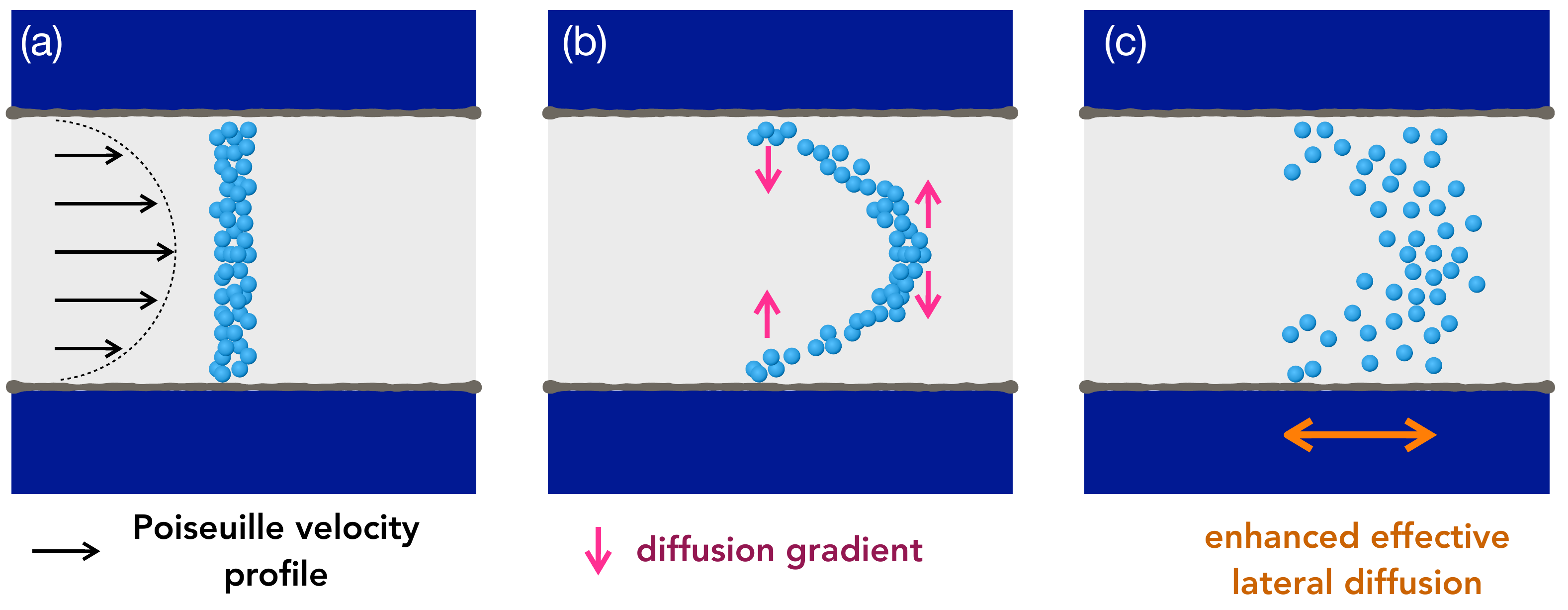}
\caption{\textbf{Taylor Dispersion in long slender channels}. Schematic illustrating how Taylor dispersion influences transport of solutes. (a) The initially radially homogeneous concentration profile is displaced by a Poiseuille flow profile. (b) This results in a strong radial and axial solute concentration imbalance, that induces radial diffusion gradients at the rear and at the front of the concentration profile. (c) After some time, the cloud of solute particles is dispersed longitudinally more than it would have by pure molecular diffusion in the absence of the flow profile.}
\label{fig:introSchem}
\end{figure}

Yet, from the fluid physics perspective dispersion of solute in straight steady channels is already intricate. As solute particles inevitably transverse between streamlines of a standard Poiseuille flow profile by molecular diffusion, the effective dispersion of solute along the channel is enhanced. Taylor and Aris provided the basis to quantify this effect~\cite{taylor1953dispersion,taylor1954dispersion,aris1956dispersion}. The cross-sectionally averaged solute concentration $C(z,t)$ along the axis $z$ of a straight channel of radius $a$ with Poiseuille profile and cross-sectional averaged flow velocity $U$ can be approximated by
 \begin{equation}
 \frac{\partial C}{\partial t} = - U  \frac{\partial C}{\partial z} +  \left( \kappa + \frac{U^2 a^2}{48 \kappa}\right)  \frac{\partial^2 C}{\partial z^2},
 \label{eq:Taylor}
 \end{equation}
where the bare molecular diffusion coefficient $\kappa$ is enhanced by a factor known as Taylor dispersion today, see Fig.~\ref{fig:introSchem}. Since the Taylor dispersion effect is inevitable in microscale devices\sop{, significant effort has been invested to understand how to control and how to model Taylor dispersion within a number of settings.}


\sop{Some of the key findings to extend Taylor dispersion in a number of situations include the understanding of the effect of reactive walls (a model for chemical reactions occurring at interfaces)~\cite{sankarasubramanian1973unsteady,lungu1982effect,smith1983effect,barton1984asymptotic,shapiro1986taylor,balakotaiah1995dispersion,biswas2007taylor,datta2008dispersion,berezhkovskii2013aris,levesque2013accounting,levesque2012taylor,levesque2013molecular},  of buoyancy effects and cross flows (to model \textit{e.g.} sedimenting particles)~\cite{erdogan1967effects,nadim1985transport,shapiro1987chemically,lin2019taylor}, of non-newtonian fluids~\cite{sankar2016mathematical} and extensions to more complex geometries such as porous media or networks~\cite{shapiro1988dispersion,brenner1980dispersion,dorfman2002generalized,dutta2001dispersion,dutta2006effect,roberts2004shear}. In driven or more complex nonequilibrium systems a number of effects still remains to be explored (\textit{albeit} the case of oscillatory shear flows as a model for the human blood vasculature~\cite{schmidt2005effect,joshi1983experimental}). For example, Taylor dispersion has entangled consequences in driven situations such as electro-osmosis or electro-phoretic transport~\cite{yariv2007electrophoretic,ng2012dispersion} and on the transport of active matter within porous media~\cite{alonso2019transport,bhattacharjee2019bacterial}. Yet, how one may control Taylor dispersion with \textit{active boundaries} is still unanswered despite its utter importance for a number of biological systems, \textit{e.g.} in situations that require mixing at very small scales~\cite{Alim:2017, cremer2016effect}, or for fabrication of advanced soft robotics. We expect dispersion to be modified by complex flows at the inlet~\cite{ng2012dispersion} or dynamically pulsating  boundaries~\cite{marbach2018transport}, but to answer how, a thorough, quantitative understanding is required.}


%



%

\sop{In this work we investigate how the effect of Taylor dispersion in straight channels can be diminished, suppressed or enhanced by specific boundary conditions: (1) \textit{controlling the flow profile entering the channel} or (2) \textit{controlling the motion of the confining walls}. To this end we employ the invariant manifold method to calculate the effective dispersion equation for the cross-sectionally averaged concentration profile~\cite{mercer1990centre,mercer1994complete}. We show that changing the flow profile at the inlet can only serve to enhance dispersion, and the optimal flow profile for mixing is only a moderate increase as compared to classical Taylor dispersion in Poiseuille flow. However, we find that space-time contracting, active channel walls open up a broad range of possibilities to control Taylor dispersion, as they give rise to corrections additive to the Taylor dispersion correction. Thereby, we show that active, slow moving, channel walls allow to suppress Taylor dispersion and overall diminish dispersion. Very fast walls allow to enhance Taylor dispersion by shuttle dispersion. We fully rationalize all these regimes and their crossover. }

\sop{Our paper is organized in the following way. In the perspective of making a pedagogical introduction to the invariant manifold method used to derive asymptotic dispersion equations for the solute concentration profile, we dedicate section II to the explanation of the method and to an example.} We revisit this rigorous method and show how versatile and powerful it can be, as it is easily applied to compute analytic results for the various settings that we explore here. Notably, we stress that while the standard textbook approach to derive Taylor dispersion may fail in more advanced settings, the invariant manifold expansion robustly yields reliable results. \sop{Section III reports the results on the impact of flow profile in straight channels, notably asking the question of how slippage modifies dispersion. Section IV reports the results on the impact of channel wall space-time contractions. We report in the Appendix the detailed calculations for each section. }


\sop{\section{Background: the invariant manifold method}}
Dispersion of solute of concentration $c(r,z,t)$ in a long, slender channel of radius $a(z,t)$ and length $L$ is fully described by the advection-diffusion equation 
\begin{equation}
\frac{\partial c}{\partial t}= -u\frac{\partial c}{\partial z}-v\frac{\partial c}{\partial r} +\kappa \frac{\partial^2 c}{\partial z^2} + \kappa \frac{1}{r}  \frac{\partial}{\partial r}  \left( r\frac{\partial c}{\partial r} \right),
\label{eq:fulladvdiff}
\end{equation}
where $u(r,z,t)$ and $v(r,z,t)$ are the axial and radial flow velocities and $\kappa$ is the molecular diffusivity. To solve Eq.~\eqref{eq:fulladvdiff} one must also specify boundary conditions for the solute flux at the channel wall. While the advection-diffusion equation is the most accurate description of dispersion, the impact that flow profiles or boundary conditions have on the effective transport and dispersion is hard to infer as two spatial components, $r$ and $z$, have to be considered. The very insightful approach by Taylor and Aris \sop{ \cite{taylor1953dispersion,taylor1954dispersion,aris1956dispersion}} has been to instead aim for a reformulation of the problem in terms of the dispersion of the cross-sectionally averaged solute concentration $C(z,t) = \frac{2}{R^2}\int c(r,z,t)\, r\, dr $, averaging out the dynamics in radial direction. While heuristically averaging out the radial direction works for dispersion in a straight channel with Poiseuille flow profile (see Appendix A) more complicated flow profiles, \textit{e.g.} dependent in $z$ and $t$, require more formal techniques. 

Among those techniques the invariant manifold method originally introduced by Mercer and Roberts \cite{mercer1994complete,mercer1990centre} is very versatile and can be readily applied to derive dispersion dynamics in long slender channels for any boundary condition and flow profile, as outlined in the following. Notably, the invariant manifold method was recently used as a rigorous proof to establish the effective diffusion expression \sop{in a straight channel}~\cite{beck2018rigorous}.  The method is based on an invariant manifold description, which has the advantage to be systematically extendable to higher orders, and thus allows to keep faithfully track of the appropriate order required and also allows to assess the magnitude of the next higher order term neglected. For more information on the invariant manifold description and other methods we refer the reader to \sop{Refs.~\cite{carr1981applications,roberts1989utility,balakotaiah1995dispersion,watt1995accurate,rosencrans1997taylor}. }

\sop{Although there exists a number of other techniques to infer the effective diffusion coefficient at long times (for example the method of moments, implicit constructions and others~\cite{gill1970exact,roberts2014model,frankel1989foundations}), the invariant manifold method is one of the only ones that can provide asymptotic dispersion equations (and not just dispersion coefficients), which is one of our aims. We refer the reader to Ref.~\cite{young1991shear} for a detailed comparison of the different methods.}

\sop{\subsection{Overview of the invariant manifold method}}

As a first step to introduce the invariant manifold method we rewrite Eq.~\eqref{eq:fulladvdiff} by introducing non-dimensional quantities: $r' = r/a$ as the radial coordinate, $z' = z/L$ the axial coordinate, $t' = t/(L/U)$ time with $U$ the mean longitudinal flow velocity in the channel, $u' = u/U$ as the longitudinal velocity, and $v' = v/(U a/L)$ the radial velocity. The advection-diffusion equation to be expanded is then given, after reordering terms, by:
\begin{equation}
\left( \frac{L}{U} \frac{\kappa}{a^2} \right) \, \frac{1}{r'}  \frac{\partial}{\partial r'}  \left( r' \frac{\partial c}{\partial r'} \right)  = \frac{\partial c}{\partial t'} +  u' \frac{\partial c}{\partial z'} +  v' \frac{\partial c}{\partial r' } - \left(\frac{L}{U} \frac{\kappa}{L^2} \right) \frac{\partial^2 c}{\partial z'^2}.
\label{eq:advDiff0}
\end{equation}
In further equations we will drop the prime notations for simplicity.
In the following we also use the abbreviated operator $\mathcal{L} c=  \frac{1}{r}  \frac{\partial}{\partial r}  \left( r \frac{\partial c}{\partial r} \right) $.

\sop{The aim is now to find an asymptotic dispersion equation. This is done typically by assuming that dynamics of solute dispersion across the channel are much faster than along the channel. More formally this means that the time to diffuse across the cross section  $\tau_{\rm diff}^a = a^2/\kappa$ is much smaller than the time to be advected along the channel $\tau_{\rm adv} = L/U$. We can therefore introduce a small parameter $\varepsilon =   \tau_{\rm diff}^a/\tau_{\rm adv}$ and look for an expansion in this small parameter~\footnote{It has further been shown that the invariant manifold method can hold even in the case where this small parameter approaches unity~\cite{rosencrans1997taylor}.}. Furthermore, we assume that the time to diffuse along the channel main axis  $\tau_{\rm diff}^L = L^2/\kappa$ is much longer than $\tau_{\rm adv}$ and that we may write $ \tau_{\rm diff}^L = \beta \varepsilon$ where $\beta$ is of order unity. The advection-diffusion equation simplifies to:
\begin{equation}
\frac{1}{\varepsilon} \, \mathcal{L} c = \frac{\partial c}{\partial t} +  u \frac{\partial c}{\partial z} +  v \frac{\partial c}{\partial r} - \varepsilon \,  \beta  \frac{\partial^2 c}{\partial z^2}.
\label{eq:advDiff}
\end{equation}

We now look for an effective equation on the cross-sectionally averaged concentration $C(z,t)$ and we aim for a systematic expansion in $\varepsilon$,}
\begin{equation}
\frac{\partial C}{\partial t} = G^1\Bigg[C,\frac{\partial C}{\partial z}\Bigg] +\varepsilon G^2\Bigg[C,\frac{\partial C}{\partial z},\frac{\partial^2 C}{\partial z^2}\Bigg] + ... = \sum_{n=1}^{\infty}  \varepsilon^{n-1} G^n\Bigg[\left\{\frac{\partial^j C}{\partial z^j}\right\},j\in [0,n] \Bigg],
\label{eq:manifoldaim}
\end{equation}
\sop{where the $G^n$ represent the different terms in the asymptotic differential equation when it is extended to order $n$ in $\varepsilon$. Note that because of the structure of the differential equation,  $G^n$ will only depend on the spatial derivatives of $C$ in $z$ of order $n$ or less. For the solute concentration we also look for a systematic expansion as }
\begin{equation}
c = V^0\Bigg[r, t; C\Bigg] + \varepsilon V^1\Bigg[r, t; C,\frac{\partial C}{\partial z}\Bigg] + \varepsilon^2 V^2\Bigg[r, t; C,\frac{\partial C}{\partial z},\frac{\partial^2 C}{\partial z^2}\Bigg] + ... = \sum_{n=0}^{\infty} \varepsilon^n V^n\left[r, t; \left\{\frac{\partial^j C}{\partial z^j}\right\},j\in [0,n]\right].
\label{eq:manifoldansatz}
\end{equation}
\sop{To infer the solution for each successive order of $V^n$ and $G^n$ we first substitute both Eqs.~\eqref{eq:manifoldaim} and~\eqref{eq:manifoldansatz} into Eq.~\eqref{eq:advDiff}. Note, that we have to be particularly careful in respecting the chain rule when taking the time derivative of $V^n$ since $V^n$ not only directly depends on time but also indirectly via it's dependence on all individual $\frac{\partial^j C(z,t)}{\partial z^j}$}:
\sop{\begin{equation}
\begin{split}
\frac{1}{\varepsilon} \mathcal{L} V^0 +  \mathcal{L} V^1 + \varepsilon \mathcal{L} V^2 \dots =& \frac{d V^0}{d t} +  u\dz{V^0} + v \frac{\partial V^0}{\partial r} -  \varepsilon \beta \frac{\partial^2 V^{0}}{\partial z^2} +\varepsilon  \frac{d V^1}{d t} + \varepsilon u\dz{V^1} +\varepsilon  v\frac{\partial V^1}{\partial r} - \varepsilon^2 \beta \frac{\partial^2 V^1}{\partial z^2} + \dots  \\
=& \frac{\partial V^0}{\partial t} + \frac{\partial V^0}{\partial C} (G^1 + \varepsilon  G^2 + ...) +  u\dz{V^0} +  v \frac{\partial V^0}{\partial r} - \varepsilon \beta \frac{\partial^2 V^{0}}{\partial z^2},  \\
&+ \varepsilon \frac{\partial V^1}{\partial t} + \varepsilon \frac{\partial V^1}{\partial C} (G^1 + \varepsilon G^2  + ...) + \varepsilon \frac{\partial V^1}{\partial C^{(1)}} \frac{\partial (G^1 + \varepsilon  G^2+ ...)}{\partial z} , \\
&+  \varepsilon u\dz{V^1} + \varepsilon  v \frac{\partial V^1}{\partial \eta} - \varepsilon^2 \beta \frac{\partial^2 V^1}{\partial z^2}  \dots \\
\end{split}
\end{equation}}
\sop{Now we may equate terms of the same order in $\varepsilon$ and obtain:}
\begin{equation}
\label{eq:mercersystem}
\begin{split}
\mathcal{L} V^0 &= 0, \, \dots \\
\mathcal{L} V^1 & = \frac{\partial V^0}{\partial t} + \frac{\partial V^0}{\partial C} G^1+  u\dz{V^0} + v \frac{\partial V^0}{\partial r}, \,  \dots  \\
\mathcal{L} V^n &= \frac{\partial V^{n-1}}{\partial t} + \sum_{\ell=1}^n \sum_{p=0}^{n-\ell}\frac{\partial V^{n-\ell}}{\partial C^{(p)}} \frac{\partial^p G^\ell}{\partial z^p}+  u\dz{V^{n-1}} + v  \frac{\partial V^{n-1}}{\partial \eta} - \beta \frac{\partial^2 V^{n-2}}{\partial z^2},
\end{split}
\end{equation}
where $C^{(p)} = \frac{\partial^p C}{\partial z^p}$. \sop{Note that $v \frac{\partial V^{n-1}}{\partial \eta}$ is of the same order as $u\dz{V^{n-1}}$, this can be seen from the fact that the divergence of the flow field is zero.} 

In addition $V^n$ has to obey a boundary condition at the channel wall. In the simplest case, we expect an impermeable wall and therefore the concentration profile has to obey a no flux boundary condition $\frac{\partial c}{\partial r}\big|_{r = 1} = 0$. This translates into
\begin{equation}
    \frac{\partial V^n}{\partial r}\bigg|_{r = 1} = 0 \, \, \mathrm{for} \, \,\, \mathrm{all} \,\, n \geq 0.
\end{equation}
Integration constants when solving the equations are further constraint as $V^n$ and its radial derivatives need to stay finite at $r=0$. Last, a simple closing equation has to be verified, namely $\int 2 V^0 r\, dr = C $ and $\int V^n r\, dr = 0$ for all $n\geq 1$. 

Solving at each order Eq.~\eqref{eq:mercersystem} and replacing the $G^n$ terms in Eq.~\eqref{eq:manifoldaim} yields the desired asymptotic dispersion equation. Finally, it is of interest to truncate the expansion at finite order (most commonly at $n=2$) to be able to use a simple analytic description of dispersion. This truncation is justified since the asymptotic series is convergent~\cite{mercer1990centre,mercer1994complete,beck2018rigorous} provided that $\varepsilon \gtrsim 1$, which corresponds to the following physical condition
\begin{equation}
\tau_{\rm adv} = L/U \,\, \gtrsim  \,\, a^2/\kappa =  \tau_{\rm diff}^a
\label{eq:validity}
\end{equation}
where $L$ is the typical length scale (along $z$) of observation, $a$ the channel radius, $\kappa$ the molecular diffusivity and $U$ the typical longitudinal velocity~\cite{mercer1994complete}. In the following we will revert to fully dimensional equations. 

Employing the invariant manifold method on the classical example of Taylor dispersion in a straight channel with steady Poiseuille flow profile yields at second order (in $\varepsilon$) the classical result of Eq.~\ref{eq:Taylor} (see Appendix B). The same result can be derived in a more heuristic way (see Appendix A). As stated earlier, more complicated dispersion dynamics are not well captured by the heuristic approach as exemplified in the following subsection.

\sop{\subsection{Example: impact of absorption at the channel wall on dispersion in a straight channel}}
The power of the invariant manifold approach versus the heuristic approach is best exemplified for a straight channel with solute absorption at the channel wall. The problem has recently been solved with the heuristic approach \cite{Meigel:2018hw}, at the same time a number of theoretical expansions and numerical solutions are available~\cite{sankarasubramanian1973unsteady,lungu1982effect,smith1983effect,barton1984asymptotic,shapiro1986taylor,balakotaiah1995dispersion,biswas2007taylor,datta2008dispersion,berezhkovskii2013aris,levesque2013accounting,levesque2012taylor,levesque2013molecular}. Among them we use Ref.~\cite{balakotaiah1995dispersion} as a benchmark for the exact expansion coefficients. 
\begin{figure}[htbp]
\center
\includegraphics[width = 0.7\textwidth]{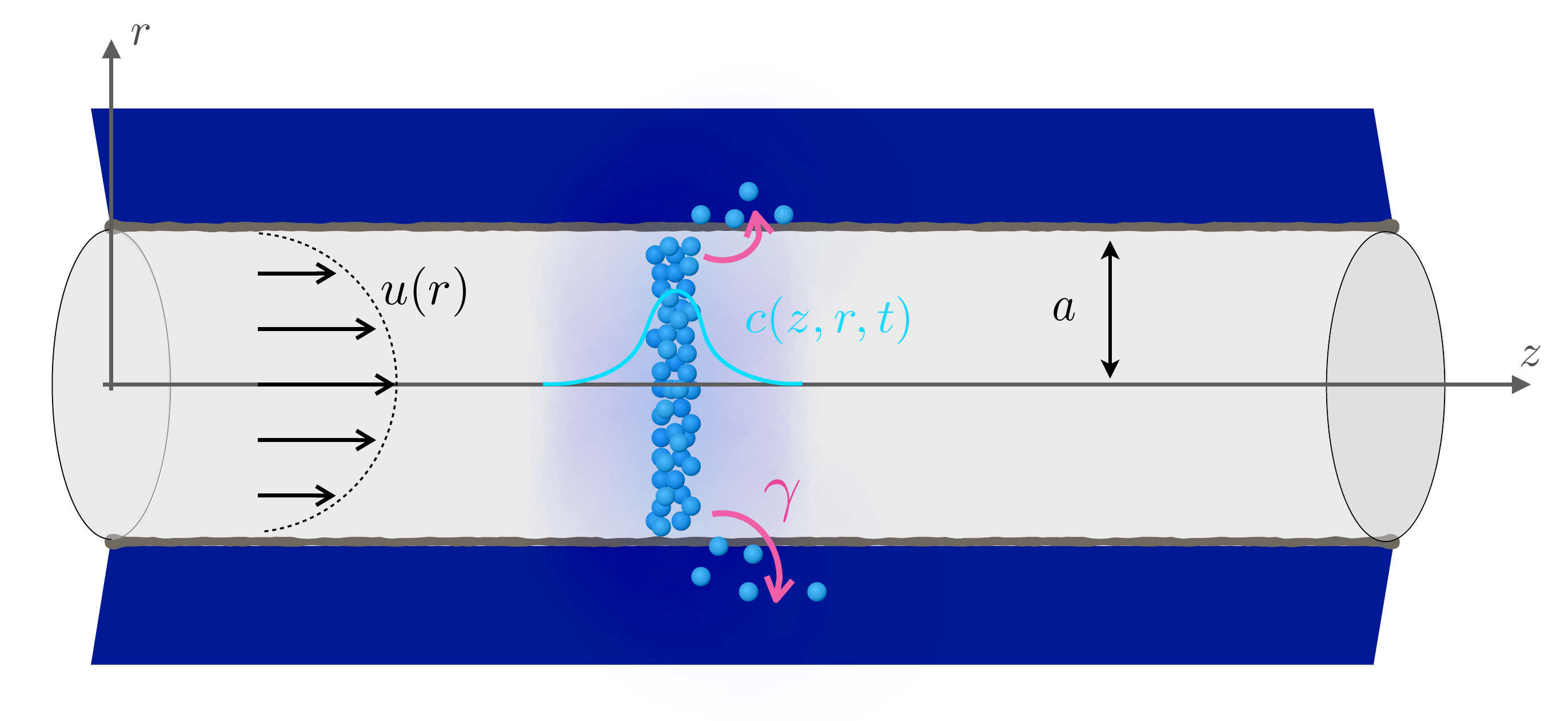}
\caption{\textbf{Setting for Taylor dispersion in a straight channel with solute absorption at the channel wall}. Straight channel schematic with solute (light blue particles) dispersing in a straight channel (light gray) and advected by a Poiseuille flow (black arrows). The diffusing particles are adsorbed with rate $\gamma$ at the channel wall.}
\label{fig:straightPipeAbsorb}
\end{figure}

The fluid is flowing in a Poiseuille profile down the channel, with zero radial flow component, $v=0$, and longitudinal flow velocity
\begin{equation*}
u=2 U\left(1-\frac{r^2}{a^2}\right),
\end{equation*}
where $U$ denotes the cross-sectionally averaged flow velocity. Solute is absorbed passively at the channel wall, see Fig.~\ref{fig:straightPipeAbsorb}:
\begin{equation*}
\kappa \frac{\partial c}{\partial r}\bigg|_{r = a} + \gamma c\big|_{r = a}= 0.
\end{equation*}
where $\gamma$ is the surface absorption coefficient in units of $\mathrm{m .\, s^{-1}}$. To solve step by step the expansion coefficients with the invariant manifold approach it is important to define how the different orders of $V^n$ are affected by the solute flux boundary condition. Since we are interested in the cross-sectionally averaged concentration $C(z,t)$ as a proxi for $c(r,z,t)$, we restrict ourselves to the case where cross-sectional gradients in absorption are averaged out quickly by cross-sectional diffusion of solute,\textit{ i.e.}~$\frac{a}{\gamma} \gg \frac{a^2}{\kappa}$, such that $\gamma a/\kappa \ll 1$. \sop{Furthermore, we assume that $\gamma a/\kappa$ is of the same order as the small number $\varepsilon$ describing the expansion of the invariant manifold method}. As a consequence, we define the boundary conditions as $\kappa \frac{\partial V^n}{\partial r}\big|_{r = a} = - \gamma V^{n-1}|_{r = a}$ for $n \geq 1$ and $\frac{\partial V^0}{\partial r}\big|_{r = a} = 0$. Working out the expansion to second order, see Appendix C, we find,
\begin{equation}
\label{eq:TD1a}
 \dt{C} = - 2\frac{\kappa}{a^2} (\gamma a/\kappa) \left( 1 -\frac{\gamma a/\kappa}{4} \right)C - \left(1 +\frac{\gamma a/\kappa}{6} \right)  U\dz{C} +  \kappa \left( 1 + \frac{a^2 U^2}{48 \kappa^2}   \right) \dzz{C}.
\end{equation}
In Eq.~\eqref{eq:TD1a}, the first term on the right hand side is a sink term corresponding to the absorption at the channel wall. The expansion allows to show that absorption increases the effective drift at second order by $U \left(1 + \frac{\gamma a/\kappa}{6}\right)$ . In fact, as particles being absorbed are closer to the walls, the particles' center of mass is effectively pushed downstream. The effective diffusion constant is exactly that of Taylor's dispersion in a straight channel, so absorption at second order does not impact diffusion here. 

\begin{figure}[htbp]
\center
\includegraphics[width = 0.95\textwidth]{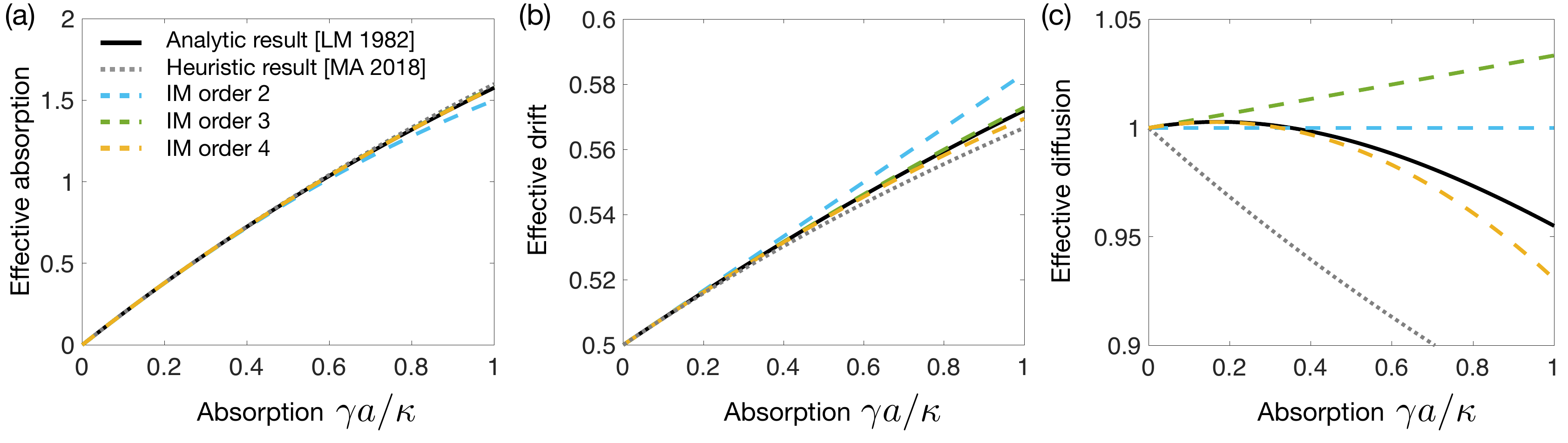}
\caption{\textbf{Coefficients of zeroth to second spatial derivatives resulting from different methods for absorption at the channel wall}. Exact coefficients [LM 1982] from Ref.~\cite{lungu1982effect} (black),
heuristic approach [MA 2018] from Ref.~\cite{Meigel:2018hw} (gray, dotted),  and invariant manifold (IM) approach (blue, green, yellow, dashed) for expansion to second, third and fourth order, respectively.
}
\label{fig:FigureCoeffs}
\end{figure}

Beyond the mechanistic understanding, it is insightful to compare Eq.~\eqref{eq:TD1a} to the equation resulting from the heuristic approach \cite{Meigel:2018hw} (or see Appendix C) and the numerical exact solutions \cite{balakotaiah1995dispersion}, see Fig.~\ref{fig:FigureCoeffs}. 
We find that the heuristic approach fails at predicting the drift and the effective diffusion accurately for moderately small absorption rates, see Fig.~\ref{fig:FigureCoeffs}. Overall the second order invariant manifold expansion, Eq.~\eqref{eq:TD1a}, only starts to show deviations from the exact results at large absorption parameters. \sop{This allows us to conclude that any heuristic derivation of slightly complex Taylor dispersion problems can be made in an unreliable way, and that more advanced techniques such as the method of moments or the invariant manifold method should be used instead.}

Note that one may push the expansion to third order, see Appendix C, to capture corrections in particular to the effective diffusion,
\begin{align}
\label{eq:TD2a}
 \dt{C} = &- 2\frac{\kappa}{a^2}(\gamma a/\kappa) \left( 1 - \frac{\gamma a/\kappa}{4}+\frac{(\gamma a/\kappa)^2}{24} \right)C -  \left(1 + \frac{\gamma a/\kappa}{6}-\frac{(\gamma a/\kappa)^2}{24}\right)  U \dz{C} \dots \\
 & + \kappa \left( 1 + \frac{a^2 U^2}{48 \kappa^2} \left[ 1+\frac{(\gamma a/\kappa)}{30}\right]  \right) \dzz{C} -\frac{a^4 U^3}{2880 \kappa^2} \frac{\partial^3 C}{\partial z^3}.
\end{align} 
 Here, we see that if absorption is small, dispersion can be enhanced by absorption. Since absorption reduces solute concentration at the channel wall, it helps to increase the radial gradient of solute concentration and thus contributes to enhancing the redistribution effect. At even higher absorption rates -- pushing the expansion to fourth order -- solute concentration is reduced near the wall too rapidly for any redistribution of solute between slow and fast streamlines to take place, thus diminishing Taylor dispersion again. The key insight gained here is that Taylor dispersion can be enhanced by changing the concentration profile of the solute across the channel's cross-section with \textit{e.g.} absorbing boundary conditions - although the magnitude of the effect is relatively small compared to pure Taylor dispersion. To investigate other means to control Taylor dispersion we now turn to the role of the flow profile in Taylor dispersion.
\section{Result 1: Impact of flow profile in a straight channel \\ on Taylor dispersion} 

\sop{The question of how to prescribe the inlet flow to control the downstream dispersion of solutes is of general interest for a number of industrial applications, and is part of the broader question of controllability of flows~\cite{glass2012prescribing,glass2016lagrangian}. }
We here investigate the impact of the axial flow profile on Taylor dispersion. We will therefore deviate from pure Poiseuille profile by first investigating the impact of slip boundaries and subsequently deriving the optimal axial flow profile for enhancing Taylor dispersion. 
\begin{figure}[htbp]
\center
\includegraphics[width = 0.7\textwidth]{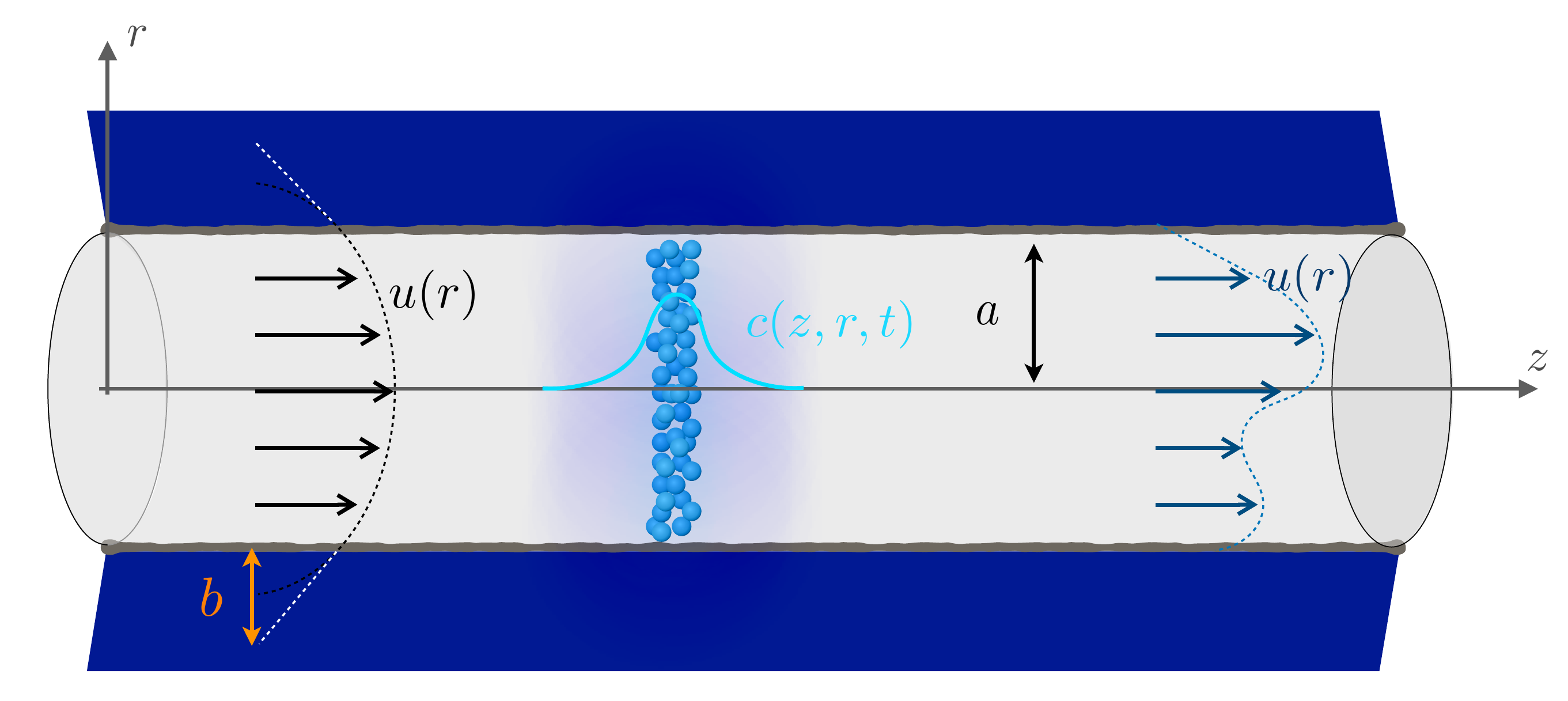}
\caption{\textbf{Setting for Taylor dispersion in a straight channel with different flow profiles}. Straight channel schematic with a solute (light blue particles) dispersing in a straight channel (light gray) and advected by a Poiseuille flow with slip at the boundaries (black arrows), with slip length $b$ (orange); or advected by an arbitrary flow profile (dark blue arrows on the right hand side). }
\label{straightPipeSlip}
\end{figure}
\subsection{Impact of slip-boundary condition at the channel wall}
In the context of microfluidics, it is interesting to focus on flow profiles that take into account different boundary conditions. In fact in micro to nano systems, the boundary properties may largely influence the flow profile in the bulk because the surface to volume ratio increases. For instance, electro-osmotic or diffusio-osmotic flows may occur, and solute concentration may also produce a feedback on the flow~\cite{tabeling2009brief}. A standard scenario arising in microfluidics~\cite{lauga2007microfluidics} and with even more impact in nanofluidics~\cite{secchi2016massive} is slip at the boundary of the channel wall.

Imposing the Navier slip boundary condition,
\begin{equation*}
\frac{\partial u}{\partial r}\bigg|_{r = a} = - \frac{a}{b}u(r=a),
\end{equation*}
where $b > 0$ is a length called the {\it slip length}, results in a vanishing radial flow velocity $v \equiv 0$ and an axial flow velocity in the channel 
\begin{equation*}
u = 2 \alpha U_0 \left( 1 - \frac{1}{\alpha} \frac{r^2}{a^2} \right),
\end{equation*}
where $\alpha = 1+2b/a$. $U_0$ is the cross-sectionally averaged axial flow velocity field in the absence of slip, characterizing a constant pressure drop through the Poiseuille law $U_0 = \frac{a^2}{8\eta} \big| \frac{\partial p}{\partial z} \big|$. Note, that the acting cross-sectionally averaged axial flow velocity $U=U_0(2 \alpha -1) = U_0\left( 1 + \frac{4b}{a} \right)$ is bigger than the flow velocity without slip.

Solving for Taylor dispersion with the invariant manifold expansion to second order, see Appendix D-1, we find,
\begin{equation}
\label{TD1b}
 \dt{C} = -U\dz{C} +  \kappa \left( 1 + \frac{a^2 U_0^2}{48 \kappa^2} \right) \dzz{C}.
\end{equation}
Eq.~\eqref{TD1b} converges to the expected solution with no slip at the boundary Eq.~\eqref{TD1}, when $b \rightarrow 0 $, as required. The effective drift that pushes the solute is indeed $U$ and not $U_0$, yet the effective diffusion coefficient is independent of the slip length. The part that counts for enhancement of dispersion is exactly the deviation of the flow profile from a plug flow. In fact, the flow velocity deviation to plug flow, $u(r) - \min_r u(r) = 2 U_0 (1 - r^2/a^2)$, corresponds to the associated Poiseuille flow with no slip at the channel wall and thus gives exactly the same dispersion enhancement as standard Poiseuille flow. In more complex situations however, coupled effects may induce a positive feedback of slippage on Taylor dispersion, \textit{e.g.} for Taylor dispersion by electro-osmotic flow that is itself increased by slippage~\cite{ng2012dispersion}.
\subsection{Impact of an arbitrary axial flow profile} 
To investigate if any axial flow profile can outcompete Taylor dispersion as arising from a Poiseuille flow profile we turn to derive Taylor dispersion for an arbitrary flow profile. \sop{This is useful to model transport through complex fluids such as Herschel–Bulkley fluids, or to perform advanced design of devices. For example one could tailor the inlet flow profile in a microfluidic device with an array of inlets with different fluid flows~\cite{lin2004generation}.} Here, the advantage of the invariant manifold method is important, in that the method is applicable independent of the characteristics of the flow profile, and thus may be used for any flow profile. Compared to similar approaches~\cite{beck2018rigorous} we find an analytic expression for the Taylor dispersion coefficient, readily applicable to any flow profile.

For a given axial flow profile $u(r)$ and zero radial flow velocity $v=0$ we find, see Appendix D-2:
\begin{equation}
\label{eq:arbitrary}
 \dt{C} = -U\dz{C} +  \kappa \left( 1 + \frac{a^2 U^2}{48 \kappa^2} I[u] \right) \dzz{C},
\end{equation}
where $U = 2 \int u(r') r' \,dr'$ denotes the cross-sectionally averaged flow velocity, with $r' = r/a$ the non-dimensional radial coordinate, and $I[u]$ is a correction factor to Taylor dispersion that is defined by
$$
I[u] = \int_0^{1} 96 r' dr' \left[ \frac{u(r')}{U} - 1\right] \left[ \frac{\tilde{u}(r')}{U} - \frac{r'^2}{4} \right], \,\,\tilde{u}(r') = \int_0^{r'} dy \int_0^{y} dx \frac{x\, u(x)}{y}.
$$
As expected, the solute is effectively advected by $U$, the cross-sectionally averaged flow velocity over the cross-section. Moreover enhancement to the effective dispersion is present via the Taylor mechanism. $I[u]$ is a measure of the strength of this correction. In this formulation it is straight-forward to see that Taylor dispersion enhancement vanishes for plug flow, \textit{e.g.}~if $u(r') \equiv U$ then immediately one finds $I[u] = 0$. If the profile is Poiseuille-like then $I[u] = 1$. It is insightful to rewrite the correction factor by introducing the non-dimensional function $\psi(r') = \frac{u(r')}{U} - 1$. As $\int_0^1r' \,dr'\, \psi(r') = 0$ this function measures the deviation of the velocity profile $u$ to a plug flow. It allows to rewrite the correction factor $I[u]$ as 
\begin{equation}
I[u] = \tilde{I}[\psi] = 96 \int_0^1 \frac{1}{x} \left[ \int_0^x x' \psi(x') dx' \right]^2.
\label{eq:enhancefac}
\end{equation}
One immediately sees that the integrand is always positive and thus there can never be solute focusing by flow in a straight channel. Any flow profile that deviates from plug flow results in $I[u] > 0$ an thus gives rise to enhanced dispersion. Yet, the remaining question is how big $I[u]$ can get.
\subsection{Optimal axial flow profile to enhance Taylor dispersion} 
\begin{figure}[htbp]
\center
\includegraphics[width = 0.7\textwidth]{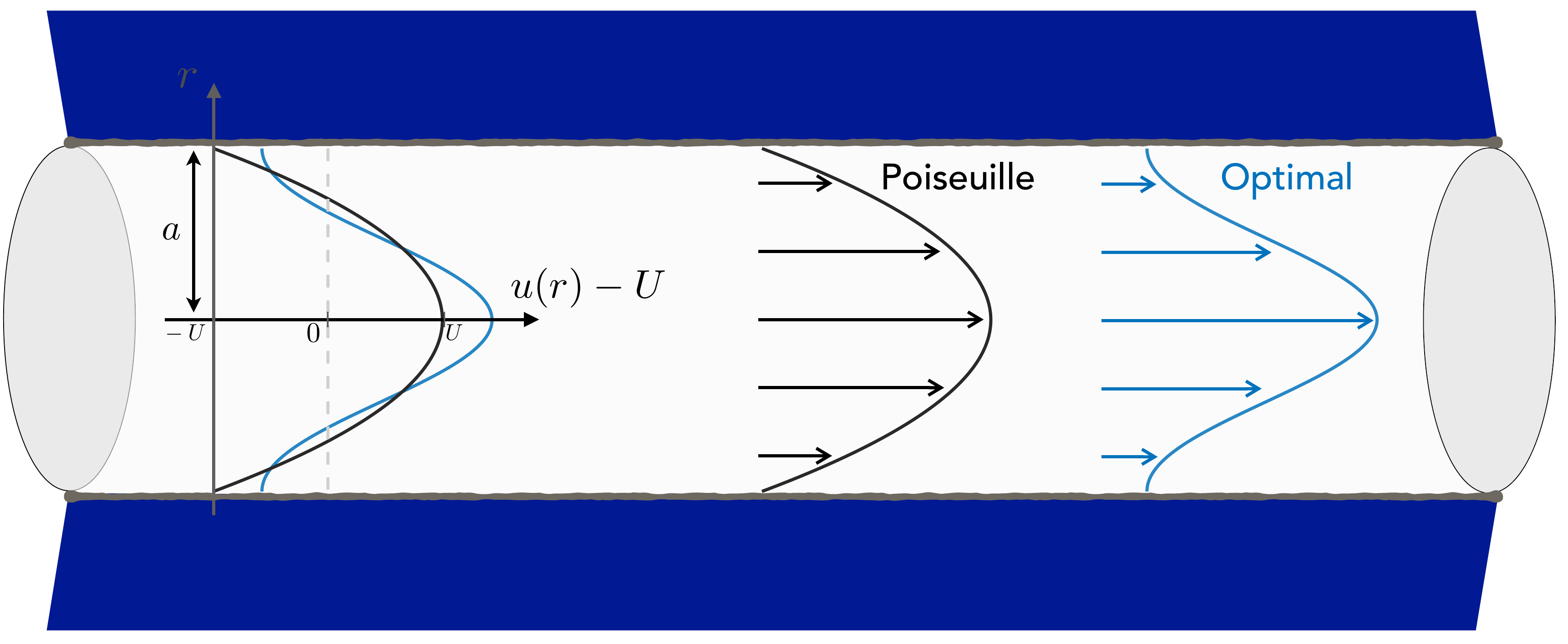}
\caption{\textbf{Optimal flow profile for Taylor dispersion in comparison with Poiseuille flow}. (right) Poiseuille profile (black) and optimal profile $u_{\rm opt}(r)$ (blue). (left) Both profiles are rescaled by the cross-sectionally averaged flow velocity $U$  to highlight their differences as a function of the radial coordinate $r$.}
\label{fig:optimalProfile}
\end{figure}

We now use the previous results (Eqs.~\eqref{eq:arbitrary} and~\eqref{eq:enhancefac}) to search for the optimal flow profile that maximizes $\tilde{I}[\psi]$ without changing the average flow velocity. Hence, we assume that the cross-sectional average of $\psi(r')$ is zero and that the deviation from plug flow has the same strength as Poiseuille flow, \textit{i.e.} the same $L^2$ norm.
Under these constraints we derive the optimal axial flow, see Appendix D-3, to be given by a Bessel function $\psi(r')=\alpha J_0\left(\frac{r'}{\sqrt{\lambda}}\right)$ with $\lambda\simeq 0.068$ and $\alpha\simeq 1.43$. Translated back to axial flow velocities this implies
$$
u_{\rm opt} (r) = U\left(1  + \alpha J_0\left(\frac{r}{a \sqrt{\lambda}}\right) \right).
$$
The optimal profile is much steeper than the Poiseuille flow, particularly with two different regions, see Fig.~\ref{fig:optimalProfile}: the central region moves really fast and the borders move slowly. This gradient will maximize the transverse diffusive drift and increase the effective lateral dispersion. That said, $I[u_{\rm opt}] \simeq 1.09$ is only slightly larger than $I[u]=1$ obtained for Poiseuille flow. Obviously, one would have to work with a very specific setting or a very peculiar fluid to obtain such a flow profile. Yet, it is fascinating to see that Poiseuille flow is almost the optimal flow profile for Taylor dispersion. In the same context it seems insightful to evaluate also the correction factor for the most common description of blood, namely a Herschel-Bulkley fluid. With $r_p = 0.2$ and $n=0.95$ (taking $r_p$ and $n$ as parameters of the Herschel-Bulkley fluid from~\cite{sankar2016mathematical}) the Taylor dispersion enhancement factor is $I[u_{\mathrm{H-B}}] \simeq 0.96$ and with $n=1.05$, $I[u_{\mathrm{H-B}}] \simeq 0.91$. Here, the flat center of the Herschel-Bulkley flow profile prevents the Taylor-dispersion effect in the center, lowering the overall enhancement of dispersion~\cite{sankar2016mathematical}. While the flow profile clearly controls Taylor dispersion, enhancing Taylor dispersion merely on the basis of the axial flow profile is limited. We, therefore, next turn to investigate the impact of active channel walls.
\section{Result 2: Impact of channel wall space-time contractions on Taylor dispersion}
\begin{figure}[htbp]
\center
\includegraphics[width = 0.7\textwidth]{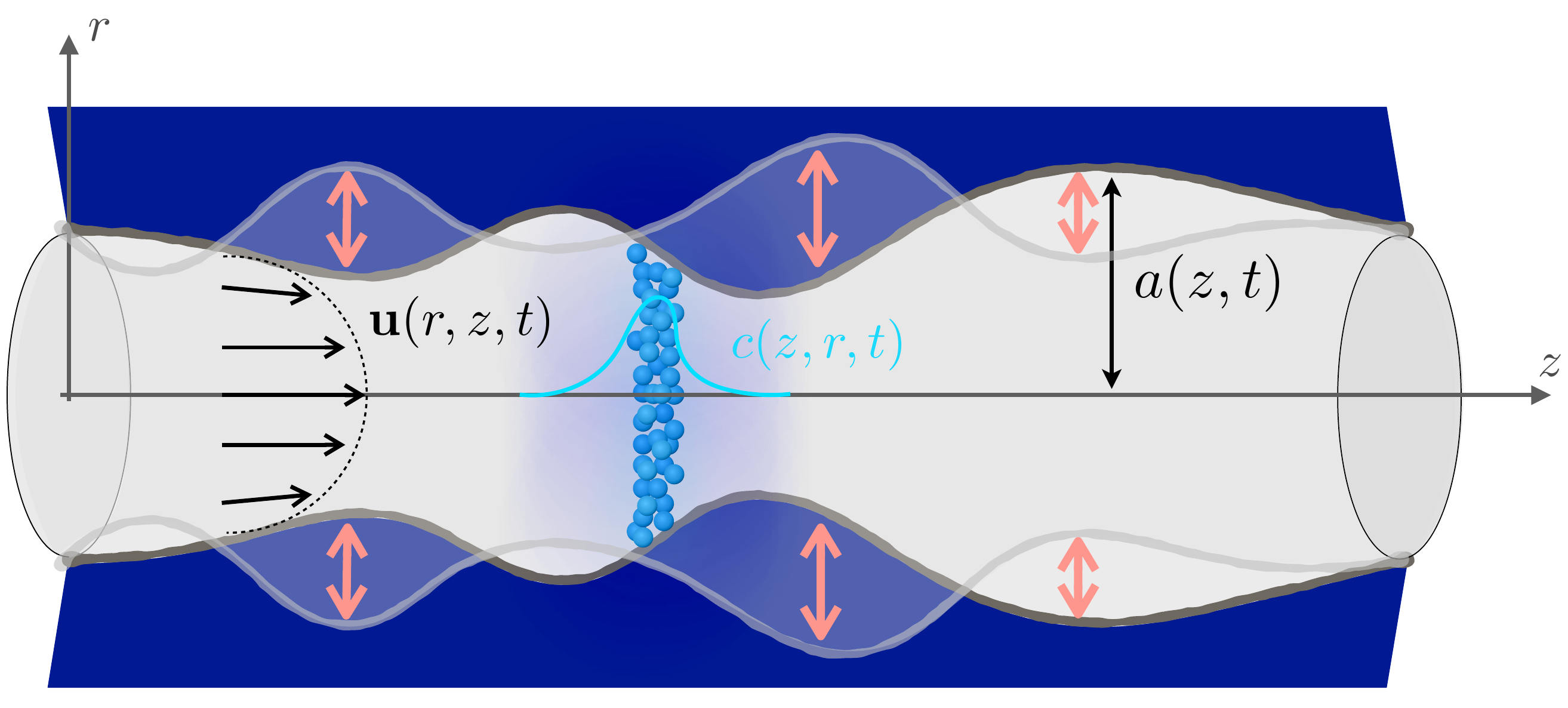}
\caption{\sop{\textbf{Setting for Taylor dispersion in a space-time contracting channel}. Channel schematic with a solute (light blue particles) dispersing in a channel (light gray) and advected by a Poiseuille like-flow (black arrows). The flow also has a radial component. The borders of the channel are actively contracting (orange arrows) and therefore moving with time.}}
\label{straightPipeSlip}
\end{figure}
Turning to active channel walls we here aim to derive Taylor dispersion in general for any space-time contraction dynamics of a channel radius $a(z,t)$ going beyond results for spatially varying channels \cite{mercer1994complete} 
and the role of oscillatory shuttle flow \cite{chatwin1975longitudinal,watson1983diffusion,leighton1988shear,schmidt2005effect,joshi1983experimental}. 
The channel contractions now imply a radial component of fluid flow due to the no-slip boundary condition at the channel wall imposing
$$
u(r=a,t)=0,\; v(r=a,t)=\frac{\partial a}{\partial t}.
$$
Taking into account the continuity equation $\frac{\partial u}{\partial z}=-\frac{1}{r}\frac{\partial}{\partial r}\left(r v\right)$ fully defines the flow velocities
\begin{eqnarray}
& \displaystyle u(r,z,t)=2U\left(1-\frac{r^2}{a^2}\right),\\
& \displaystyle v(r,z,t)=\frac{\partial a}{\partial t}\frac{r}{a}(2-\frac{r^2}{a^2})+2U\frac{\partial a}{\partial z}\frac{r}{a}(1- \frac{r^2}{a^2}),
\end{eqnarray}
where the cross-sectionally averaged axial flow velocity $U$ follows from conservation of mass $\frac{\partial }{\partial z} \left( a^2 U\right) = - \frac{\partial}{\partial t} a^2$, resulting in
\begin{equation}
U(z,t) = U(z=0,t)-\frac{1}{a^2}\int_0^zd\tilde{z}\frac{\partial a^2(\tilde{z},t)}{\partial t}.
\label{eq:flowcons}
\end{equation}
\subsection{Long time concentration evolution in a space-time contracting channel}
Using  the invariant manifold method up to second order we find for the dynamics of the cross-sectionally averaged concentration $C$ (see Appendix E),
\begin{equation}
 \dt{a^2C} = \frac{\partial}{\partial z} \left( - U a^2C +  a^2 \kappa \left( 1 + \frac{a^2 U^2}{48 \kappa^2} \right) \dz{C} \right).
\label{eqMass}
\end{equation}
 Eq.~\eqref{eqMass} is eventually remarkably simple and resembles the classical Taylor Dispersion equation in the case of a steady straight channel Eq.~\eqref{eq:Taylor}. With the notation of  Eq.~\eqref{eq:Taylor}  we may identify as the relevant variable, $\rho=a^2 C$ the particle mass in an infinitesimal slice (equivalent to the marginal probability to find solute particles at coordinate $z$ and $t$). In Eq.~\eqref{eqMass} particle mass is advected with the effective drift $U$, the cross-sectionally averaged axial flow velocity, while effective diffusion is proportional to the gradient of the solute itself $\dz{C}$.  \sop{One may readily use Eq.~\eqref{eqMass} to model solute dispersion in dynamic channels (or networks of channels), not limited to but particularly prevalent, in living systems.}

 Rewriting one more time in terms of particle mass $\rho$ only, we find 
\begin{equation}
 \dt{\rho} = \frac{\partial}{\partial z} \left(-\left[ U + \frac{\kappa}{a^2}\dz{a^2} \left( 1 + \frac{a^2 U^2}{48 \kappa^2} \right)\right]\rho+ \kappa \left( 1 + \frac{a^2 U^2}{48 \kappa^2} \right) \dz{\rho}\right).
\label{eq:eqMass}
\end{equation}
In this formulation we see that the particle mass's drift is modulated by the deformation of the channel $\dz{a^2}$. For a pure Brownian particle in a channel in the absence of any flow the impact of confinement shape on a particle's effective's drift is known as entropic slow down \cite{Martens:2013}. 
Here, we find that flow velocity modulates entropic slow down and may augment the effect. Yet, how strongly the Taylor dispersion term $\frac{a^2 U^2}{48\kappa}$ changes effective drift and effective dispersion cannot be estimated in this very general form, since here both $a(z,t)$ and $U(z,t)$ depend on space and time. We therefore turn to the example of peristaltic pumping to estimate the impact of all terms on effective dispersion. 
\subsection{The impact of peristaltic pumping on dispersion with emphasis on the role of Taylor dispersion}
To estimate the impact of peristaltic pumping on Taylor dispersion we now consider explicit space-time contraction of the channel wall of the functional form
\begin{equation}
a(z,t)^2 = a_0^2 ( 1 + \phi \cos (\omega t - k z)). 
\end{equation}
Note, that to first order in $\phi\ll 1$ this ansatz is equivalent to the classical example of peristaltic pumping \cite{shapiro1969peristaltic}, yet this form is analytically more tractable \cite{Alim:2013}. From conservation of mass, Eq.~\eqref{eq:flowcons}, it follows that 
\begin{equation}
a^2(z,t) U(z,t)= a^2(z=0,t) U(z=0,t)-\omega a_0^2 \frac{\phi}{k}[\cos(\omega t-k z)-\cos(\omega t)].
\end{equation}
To allow a fully analytic calculation to gain insight into the process, we need to simplify the space-time dependence of the mean flow $U$. To this end, we consider an oscillatory inflow boundary condition $Q(z=0,t) = a^2 U(z=0,t)  = -\omega a_0^2 \frac{\phi}{k}\cos(\omega t)$ accounting for the flow generated elsewhere in a long tube. With this assumption we find
\begin{equation}
U = \frac{\phi \omega}{k} \cos(\omega t - k z)\left(1 - \phi \cos(\omega t - k z)\right) + \mathcal{O}(\phi^2),
\end{equation}
where we used the fact that $\phi \ll 1$. We further shift our coordinate system by performing the change of variables: $\xi \leftarrow \frac{\omega}{k}t - z$  and $t \leftarrow t$ such that Eq.~\eqref{eq:eqMass} becomes a dissociable function of $\xi$ and $t$ as
\begin{equation}
\begin{split}
& \dt{\rho}  = \frac{\partial}{\partial \xi} \left(-\tilde{U}(\xi)\rho + \tilde{\kappa}(\xi)  \frac{\partial\rho}{\partial \xi}\right),\\
\mathrm{with}  \,\, \tilde{U}(\xi) =  \frac{\omega}{k} - U + &\frac{\kappa}{a^2}\frac{\partial a^2}{\partial \xi} \left( 1 + \frac{a^2 U^2}{48 \kappa^2} \right) \,\,  \mathrm{and} \, \, \tilde{\kappa}(\xi) = \kappa \left( 1 + \frac{a^2 U^2}{48 \kappa^2} \right),
\end{split}
\label{eq:advDiffDG}
\end{equation}
where now advection and drift terms all depend only on $\xi$. Eq.~\eqref{eq:advDiffDG} is now amenable to further analytic computations. Either based on the calculation of mean first passage times~\cite{reimann2001giant} or on the analysis of  probability distribution functions~\cite{guerin2015kubo} it is possible to derive the effective dispersion coefficient at infinitely long times, see Appendix F, arriving at, 
\begin{equation}
\kappa_{\mathrm{eff}} = \kappa\left( 1 + \frac{Pe^2}{48} + \frac{1}{2}\phi^2\frac{6 Pe^2}{2Pe^2+(k\phi a_0)^2}-\frac{1}{2}\phi^2\frac{(k\phi a_0)^2}{2 Pe^2+(k\phi a_0)^2} \right),
\label{eq:keff1}
\end{equation}
where we here adopted the Taylor dispersion notation in terms of the P\'eclet number. To allow comparison to the case of the straight steady channel, we here define a time averaged P\'eclet number as $Pe=\sqrt{\frac{\langle U^2 a^2\rangle_{T}}{\kappa^2}}=\frac{\phi \omega a_0}{\sqrt{2} k \kappa}$, where $\langle . \rangle_T$ indicates the time-average over a contraction period $T = 2\pi/\omega$. We identify two correction terms to the Taylor dispersion term in $Pe^2/48$. 

The first correction,  $\frac{1}{2}\phi^2\frac{6 Pe^2}{2Pe^2+(k\phi a_0)^2}$, is a shear dispersion contribution. It arises as the alternating shear pushes particles back and forth, and as they diffuse on top of that, effectively increasing effective dispersion. This effect was identified separately in the case of periodic shuttle flows in straight channels~\cite{chatwin1975longitudinal,watson1983diffusion,leighton1988shear,schmidt2005effect,joshi1983experimental} and is sometimes referred to as shear dispersion just like Taylor dispersion. We must stress that the two effects are entirely different: Taylor dispersion arises because of shear in the lateral direction, so because of a non-uniform flow profile in the lateral direction but arises independently of dispersion enhanced by shuttle streaming. We therefore refer here to the later effect as \textit{shuttle dispersion}.

The second correction, $-\frac{1}{2}\phi^2\frac{(k\phi a_0)^2}{2 Pe^2+(k\phi a_0)^2}$ is a negative contribution to effective dispersion. It occurs as the channel walls are oscillating in space, therefore creating constrictions that are hard to transverse and cavities that act as particle traps. This effect counteracts dispersion, a phenomenon called \textit{entropic slow down}. It was previously identified in the absence of temporal variations, or more formally when $Pe = 0$, by numerous authors~\cite{burada2009entropic,Martens:2013,malgaretti2013entropic,yang2017hydrodynamic} and is here generalized for additional flow. Remarkably, all these contributions to effective dispersion are additive, which is not necessarily obvious. This is particularly striking since all of these effects are different in their origin and may arise independently. They are now easily comparable through Eq.~\eqref{eq:keff1}. 
\begin{figure}[htbp]
\center
\includegraphics[width = 0.8\textwidth]{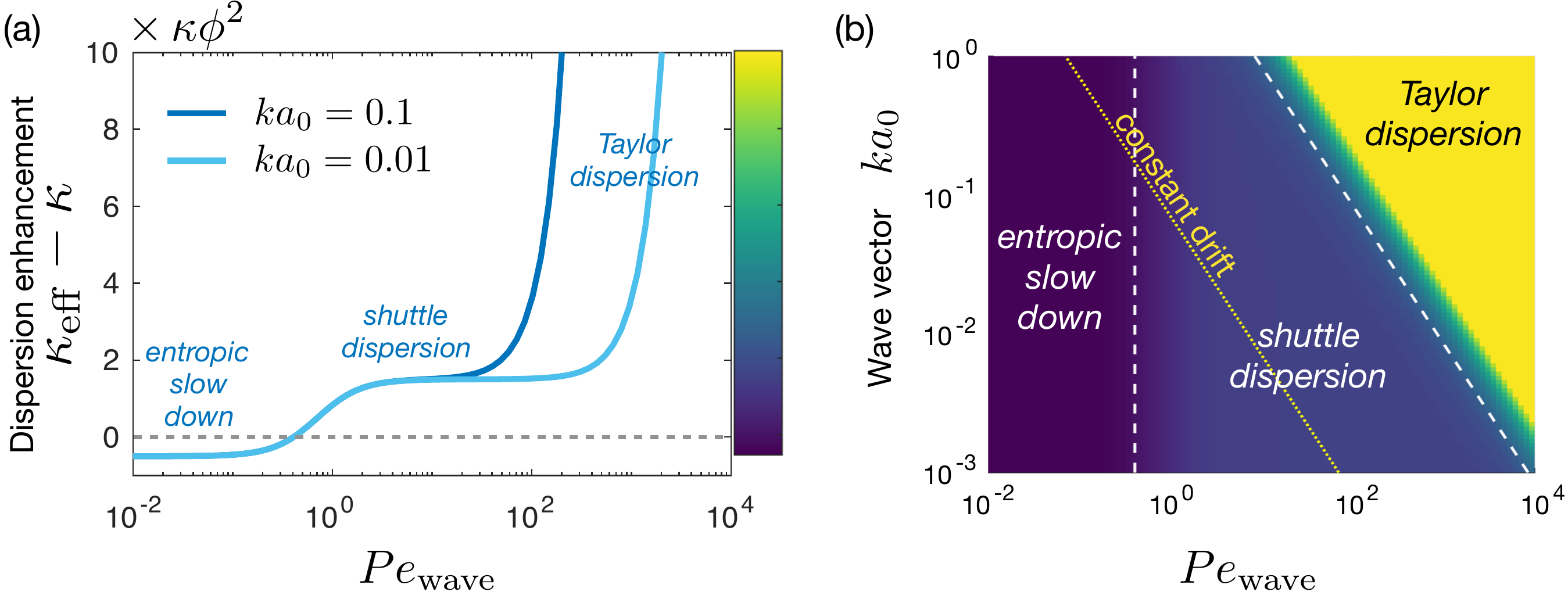}
\caption{\sop{\textbf{The different regimes of dispersion enhancement in a peristaltic channel}. (a) Dispersion enhancement defined $\kappa_{\rm eff} - \kappa$ with respect to the P\'{e}clet of the wave $Pe_{\rm wave} = \omega/(\sqrt{2} \kappa k^2)$ for different values of the wave number $k$ and (b) associated phase map in the $(ka_0,Pe_{\rm wave})$ plane of the dispersion enhancement. The color scale corresponds to the colorbar of (a) saturated at $10$ for readability. The line separating the entropic slow down and the shuttle dispersion regions corresponds to $Pe_{\rm wave}^2 = 1/6$ and the line separating the Taylor dispersion region to $(ka_0)^2 = 3\times 48/(2 Pe_{\rm wave}^2 +1)$ or nearly $ka_0 \simeq 8/Pe_{\rm wave}$. The dotted yellow line of constant drift corresponds $v_{\rm eff}$ constant, \textit{i.e.} to $Pe_{\rm wave}  = (0.1 / \sqrt{2}) \times 1 /k a_0$. All lines of constant drift are parallel to this one, with increasing velocities going towards the right.}}
\label{phaseMap}
\end{figure}

To estimate the magnitude of the correction terms relative to the Taylor dispersion coefficient in a straight channel it is insightful to rewrite the effective dispersion in terms of a P\'eclet number for advective versus diffusive transport along the contraction wave of wave number $k$, $Pe_{\mathrm{wave}}=Pe/(k\phi a_0)=\omega/(\sqrt{2}\kappa k^2)$,
\begin{equation}
\kappa_{\mathrm{eff}} = \kappa\left( 1 + \phi^2 (k a_0)^2\frac{Pe_{\mathrm{wave}}^2}{48} + \frac{1}{2}\phi^2\frac{6 Pe_{\mathrm{wave}}^2}{2Pe_{\mathrm{wave}}^2+1}-\frac{1}{2}\phi^2\frac{1}{2 Pe_{\mathrm{wave}}^2+1} \right).
\end{equation}
In this notation we see that all corrections to molecular diffusivity $\kappa$ scale with $\phi^2$, yet the Taylor dispersion correction has an additional prefactor $(k a_0)^2$ which is always smaller than one and particularly small for long wavelengths of the space-time contractions. Exploring the limits, we find that for $Pe_{\mathrm{wave}}\ll 1$ -- a slowly contracting channel --, shuttle dispersion and Taylor dispersion corrections vanish, with Taylor dispersion vanishing faster. Entropic slow down dominates and simplifies to $-\frac{1}{2}\phi^2$. For more intermediate values of $Pe_{\mathrm{wave}}$ entropic slow down vanishes and shuttle dispersion simplifies to $\frac{3}{2}\phi^2$, while Taylor dispersion may not yet be relevant. For very large $Pe_{\mathrm{wave}}$ however, the Taylor dispersion correction again dominates and continues to increase with $Pe_{\mathrm{wave}}$ as $Pe_{\mathrm{wave}}^2$. The different regimes are illustrated in Fig.~\ref{phaseMap}, that clearly demonstrates that different parameters will yield very different corrections to dispersion. Although the transition between entropic slow down and shuttle dispersion was clearly evidenced by one of the authors~\cite{marbach2018transport}, it is now striking to see how Taylor dispersion dominates either regime in a range of parameters, typically when $ka_0 \gtrsim 8/Pe_{\rm wave}$ or more simply when $Pe \gtrsim 8$. Note, this requirement is also compatible with the validity condition of the framework Eq.~\eqref{eq:validity} as long as $L/a_0 \gg 1$. The results displayed in Fig.~\ref{phaseMap} suggest a number of ways to control and tune dispersion in an active channel. 

\sop{With such peristaltic pumping, one finds also that the fluid within the channel is advected with a mean non-zero effective drift (see Appendix F) that writes 
\begin{equation}
v_{\rm eff} =\frac{\omega}{k} \frac{\phi^2}{2}. 
\end{equation}
When the contractions disappear, naturally this effective drift disappears. Making the contractions bigger (\textit{i.e.} $\phi$ bigger) increases the effective drift (by increasing the displaced mass). The effective drift is in the direction of the peristaltic wave and scales with the phase velocity $\omega/k$. One may therefore control the direction and the amplitude of the effective drift by tuning the parameters of the peristaltic pump. One may also keep the effective drift constant while changing the effective dispersion (see Fig.~\ref{phaseMap}-b). Note that one may change further this effective velocity by adapting the boundary conditions at the inlet and outlet of the tube. } 

\section{Discussion and Conclusion}
The recent shift to active and smart artificial systems requires to revisit Taylor \sop{dispersion} afresh and explore the impact of specific boundary conditions on the walls and extend to active channel walls. Moreover, we now have the ability to design innovative systems and to make refined observations with emerging techniques that allow us to quantify flows at very small scales in various biological contexts -- flows generated by acto-myosin cortical activity in amoeba~\cite{lewis2015coordination}, slime molds~\cite{matsumoto2008locomotive} or plant cells~\cite{peremyslov2015myosin} or flows generated by cilia motion~\cite{jonas2011microfluidic} -- but also in nanoscale artificial systems down to a few fL/s~\cite{secchi2016massive}. Therefore, the ability to model how specific flows and boundary conditions affect solute transport within channels is critical to understand solute transport through these systems.

\sop{By revisiting a rigorous and systematic method, the invariant manifold approach, we showed that we could assess the impact of various boundary conditions (solute absorption, flow slippage), flow profile and active channel walls on long time advection and dispersion of the solute. We stress here again that the invariant manifold approach provides a rigorous framework that allows to derive an effective advection-diffusion equation for the solute, along the principal axis of interest (the long axis of a tube for example) in a systematic way. Importantly, this method avoids averaging mistakes in more complex situations than just a straight tube (as for absorption boundary conditions for example), and also allows to build more refined approximations with \textit{e.g.} higher order derivatives in the effective advection-diffusion equation.}

\sop{We recapitulate the specific results of our work:}

\vspace{1mm}

\noindent \sop{(1) \textit{controlling the flow profile entering the channel} gives a small handle to tune the effective dispersion. In particular, we have shown that allowing for slip on the channel walls, but still with a constant pressure drop along the channel does not modify dispersion. Furthermore, the Poiseuille flow profile in a cylindrical channel is nearly optimal for dispersion. As flow profiles get \enquote{flatter}, the dispersion enhancement diminishes to zero (zero being obtained for a plug flow profile). }

\vspace{1mm}

\noindent \sop{(2) \textit{controlling the motion of the confining walls} gives a huge handle to tune both effective dispersion and effective drift of the solute. The average solute concentration profile $C(z,t)$ at location $z$ on the principal axis of the tube may be described by
\begin{equation}
 \dt{a^2C} = \frac{\partial}{\partial z} \left( - U a^2C +  a^2 \kappa \left( 1 + \frac{a^2 U^2}{48 \kappa^2} \right) \dz{C} \right).
\label{eqMassEnd}
\end{equation}
where $a(z,t)$ is the radius of the pulsating confining wall and $U$ is the mean fluid velocity triggered by the contractions and flow incompressibility. This effective model for the concentration profile evolution is valid as long as Eq.~\eqref{eq:validity} is verified, that we recall here for consistency:
$\frac{L}{U} \gtrsim \frac{a^2}{\kappa}$.    
Note that it is possible to go beyond this regime, by pursuing the expansion and allowing for higher order derivatives in Eq.~\ref{eqMassEnd}.
Furthermore, for active channel walls, we showed that it is possible to tune dispersion between three different regimes of dispersion arising in a parameter space with two degrees of freedom: the contraction frequency and the contraction wavelength of the active channel walls.}

\vspace{1mm}

Our work highlights the clear versatility of the invariant manifold method. It would be insightful to see the method be used to investigate further settings beyond the scope of this work for example coupling various effects, and to obtain also simple and readily applicable results for instance to pinpoint the role of diffusio-osmotic or electro-osmotic flows on dispersion~\cite{ng2012dispersion} with different boundary conditions. 

\sop{Beyond the regimes that we have considered -- where our continuum model is applicable -- a number of extensions are possible. For example, it would be enlightening to understand how large peristaltic contractions affect the effective dispersion (with $\ell \simeq a$ where $\ell$ is the typical wavelength of the contraction). This could be accessible via continuum simulations with realistic membrane models for the tubes~\cite{deserno2015fluid}. Furthermore, when the system size drops to nanoscales, continuum theories will fail to describe in a reliable way the velocity field in the fluid, and other effects. Only brownian dynamics or molecular dynamics are able to access these scales, assess the behaviors at play~\cite{yoshida2018dripplons} and inform us on the possible mechanisms by which dispersion may be enhanced or decreased. More efforts are needed in particular to bridge the gap between those discrete scales and continuum scales.}

\begin{acknowledgments}
The authors are indebted to Lyd\'{e}ric Bocquet, Agnese Codutti, Michael P.~Brenner, David S. Dean, Fr\'{e}d\'{e}ric Marbach and Felix J.~Meigel for fruitful discussions. S.~M.~acknowleges funding from A.N.R. Neptune. K.~A.~acknowledges funding from the Max Planck Society. 
\end{acknowledgments}
\section*{Appendix}
\subsection{Textbook derivation of Taylor dispersion}
\label{sec:textbook}
\begin{figure}[h!]
\center
\includegraphics[width = 0.7\textwidth]{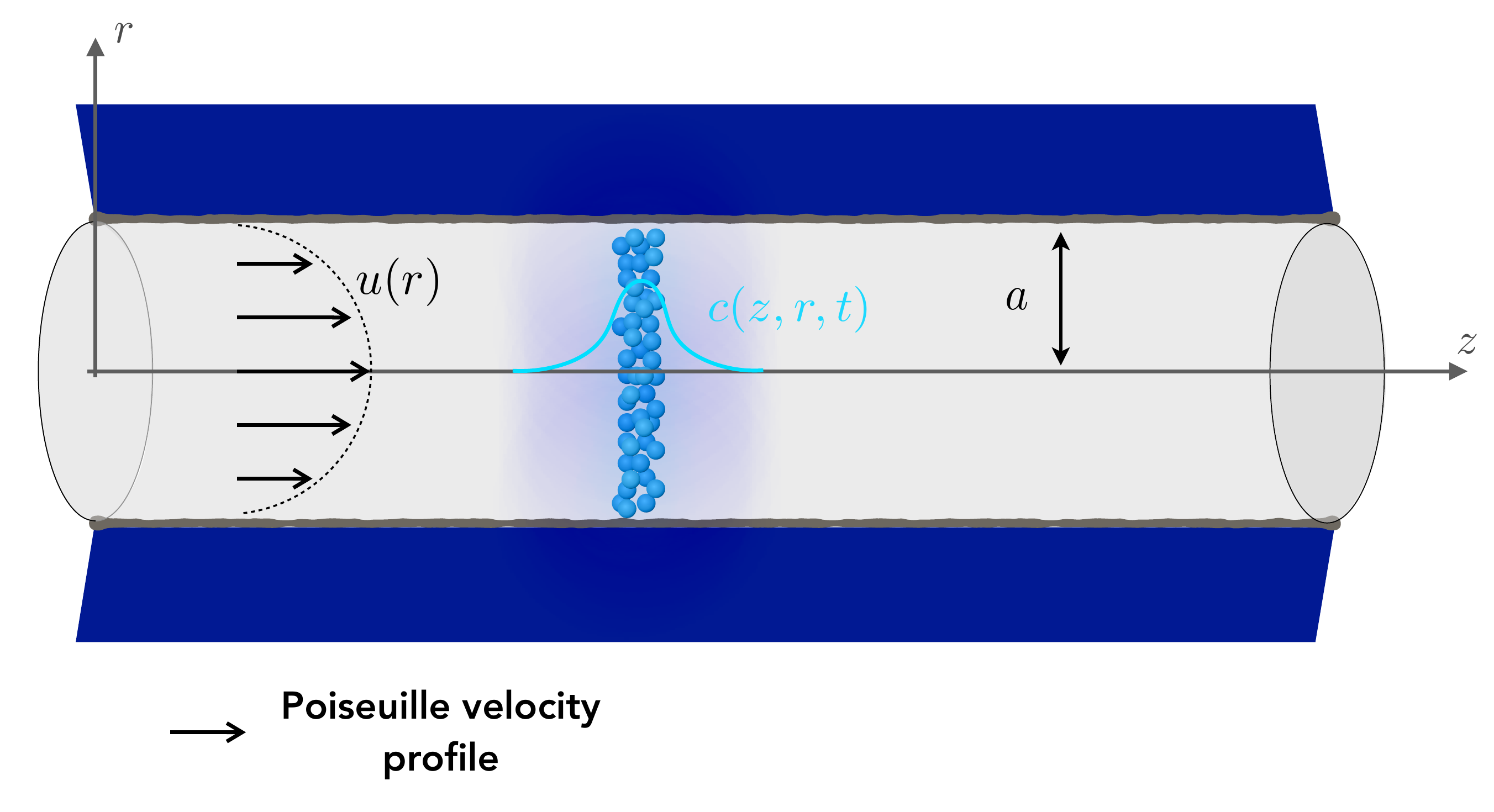}
\caption{\textbf{Setting for Taylor dispersion in a straight channel}. Straight channel schematic with a solute (light blue particles) dispersing in a straight channel (light gray) and advected by a Poiseuille flow (black arrows). }
\label{straightPipe}
\end{figure}

Here we revisit for completeness Taylor dispersion as it is typically introduced in a textbook (e.g.~\cite{bruus2008theoretical}). Consider solute dispersing in a straight cylindrical channel of radius $a$ with radial and longitudinal coordinates $(r,z)$, see Fig.~\ref{straightPipe}. 
At low Reynolds numbers, the flow field induced by a pressure drop in the channel obeys a Poiseuille profile 
$$
u(r) = 2 U \left(1- \left(\frac{r}{a}\right)^2\right),
$$
where $U$ is the cross-sectional average of the flow velocity $u(r)$. Apart from this definition of $U$ we will use an overline $\overline{f}$ to denote the cross-sectional average of a variable $f$ in the following.  Suppose at time $t=0$, an axisymmetric solute concentration is released $c(z,r,t)$. Then the solute concentration obeys the following advection diffusion equation:
\begin{equation}
\label{textbookDiff}
\frac{\partial c}{\partial t} + u(r) \frac{\partial c}{\partial z} = \kappa \left(\frac{\partial^2 c}{\partial z^2}+\frac{1}{r} \frac{\partial}{\partial r}\left(r \frac{\partial c}{\partial r}\right) \right)\,\, \mathrm{with} \,\, \frac{\partial c}{\partial r}\bigg|_{r=a} = 0.
\end{equation}
We look for an equation for $C(z,t) = \overline{c(z,r,t)}$ assuming that $\tilde{c}(z,r,t) = c(z,r,t) - C(z,t)$ is a small correction to the cross-sectional average concentration. For coherence with the following section we also write $C(z,t)= C(z,t)$. Inserting that in Eq.~\eqref{textbookDiff} gives
\begin{equation}
\label{textbookDiffav}
\frac{\partial C}{\partial t} +\frac{\partial \tilde{c}}{\partial t} + u(r) \frac{\partial C}{\partial z}+ u(r) \frac{\partial \tilde{c}}{\partial z}  = \kappa \left(\frac{\partial^2 C}{\partial z^2}+\frac{\partial^2 \tilde{c}}{\partial z^2}+\frac{1}{r} \frac{\partial}{\partial r}\left(r \frac{\partial \tilde{c}}{\partial r}\right) \right),
\end{equation}
and taking the cross-sectional average and using the boundary conditions gives
\begin{equation}
\label{textbookDiffav2}
\frac{\partial C}{\partial t} + U \frac{\partial C}{\partial z} + \overline{u(r) \frac{\partial \tilde{c}}{\partial z}} = \kappa \frac{\partial^2 C}{\partial z^2}.
\end{equation}
Eq.~\eqref{textbookDiffav2} is the partial differential equation for $C$ that we are looking for. The problem is to evaluate $ \overline{u(r) \frac{\partial \tilde{c}}{\partial z}}$. Subtracting Eq.~\eqref{textbookDiffav2} from Eq.~\eqref{textbookDiffav} gives:
\begin{equation}
\label{textbookcp}
\frac{\partial \tilde{c}}{\partial t}  + \left(u(r) - U\right)\frac{\partial C}{\partial z}+ \left(u(r) \frac{\partial \tilde{c}}{\partial z} - \overline{u(r) \frac{\partial \tilde{c}}{\partial z}} \right)= \kappa \left(\frac{\partial^2 \tilde{c}}{\partial z^2}+ \frac{1}{r} \frac{\partial}{\partial r}\left(r \frac{\partial \tilde{c}}{\partial r}\right) \right).
\end{equation}
We can use this expression to solve for the missing term but we have to perform coarse grain approximations. First we look for \enquote{long time} solutions, typically for times longer than the typical time to diffuse across the cross-section $t \gg \frac{a^2}{\kappa}$, where we expect that the cross channel diffusion will have averaged out such that we may neglect $ \tilde{c}_t $. For such time scales we also expect $\tilde{c} \ll C$, such that we may also neglect $\left(u(r) \frac{\partial \tilde{c}}{\partial z} - \overline{u(r) \frac{\partial \tilde{c}}{\partial z}} \right)$. Furthermore, since $\tilde{c}$ is essentially a radial correction we assume that gradients in the $r$-direction for $\tilde{c}$ will be much greater than those in the $z$-direction. As a result Eq.~\eqref{textbookcp} reduces to:
\begin{equation}
\left(u(r) - U\right) \frac{\partial C}{\partial z} = \kappa \frac{1}{r} \frac{\partial}{\partial r}\left(r \frac{\partial \tilde{c}}{\partial r}\right) .
\end{equation}
Integrating, using that $\tilde{c}$ is finite at $r=0$ and the fact that the cross-sectional average of $\tilde{c}$ should vanish yields:
\begin{equation}
\tilde{c}(z,r,t) = \frac{a^2 U}{24 \kappa}  \frac{\partial C}{\partial z}\left( 6 \frac{r^2}{a^2} - 3 \frac{r^4}{a^4}    - 2 \right).
\end{equation}
Then one is left to perform the cross-section average of $u(r) \frac{\partial \tilde{c}}{\partial z}$ and one finds rather easily that Eq.~\eqref{textbookDiffav2} can be written as:
\begin{equation}
\frac{\partial C}{\partial t} = - U  \frac{\partial C}{\partial z} +  \left( \kappa + \frac{U^2 a^2}{48 \kappa}\right)  \frac{\partial^2 C}{\partial z^2}  .
\label{taylorTextbook}
\end{equation}
This equation is called the Taylor dispersion equation. We notice that it is a simple advection diffusion equation along the axial direction, with an effective drift that is exactly the average drift, and an effective diffusion that is enhanced as compared to the bare diffusion $\kappa$. This enhancement factor is due to the fact that the Poiseuille flow is not a plug flow, and thus advects particles at different radial positions with different speeds. After a typical time $\tau$ the displacement will be $\delta \ell = U\tau$. This induces a radial and axial shift in the dispersion profile, that generates radial diffusion gradients. To equilibrate over the cross section, the time $\tau$ has to match the time $a^2/\kappa$. As a result the effective diffusion due to the Poiseuille flow is $\delta \ell ^2 / \tau = \frac{U^2a^2}{\kappa}$, which is exactly the enhancement factor deduced analytically in Eq.~\eqref{taylorTextbook}. 
\subsection{Invariant manifold method for classical Taylor dispersion in a straight channel}
\label{sec:Manifoldtextbook}
We here derive Taylor dispersion again for a straight channel with Poiseuille flow profile, but now with the invariant manifold method. $U$ stands for the cross-sectional average of $u$, and the fluid flow reads:
\begin{equation*}
\begin{cases}
u = 2 U (1- \eta^2) \\
v = 0
\end{cases}
\end{equation*}
where now $\eta = r/a$ is the non-dimensional radial coordinate (to lighten notations compared to $r'$ in the main text), and the only non-dimensional variable that we keep for the derivation. 
\paragraph{0th order expansion}
We first solve for $V^0$. 
$\mathcal{L} V^0 = 0$ with $\dev{V^0}\big|_{\eta = 1} = 0$ gives $V^0 = C$. Note, that $V^0$ does not depend explicitly on $t$ nor $\eta$, therefore $\frac{\partial V^0}{\partial t}=0$. 
\paragraph{1st order expansion}
Now we solve for $V^1$. The defining equations are:
\begin{equation*}
\begin{cases}
\frac{\kappa}{a^2 \eta} \de \left( \eta \dev{V^1} \right) = \dt{V^0} +  \dC{V^0}G^1 + u\dz{V^0}, \\
\dev{V^1}\big|_{\eta = 1} = 0.
\end{cases}
\end{equation*}
We use the expression for $u(\eta)$ and integrate once with respect to $\eta$:
\begin{equation*}
\eta \dev{V^1} = \frac{a^2}{\kappa} \left( \frac{\eta^2}{2} G^1 + 2U \left[ \frac{\eta^2}{2} - \frac{\eta^4}{4} \right] \frac{\partial C}{\partial z} \right) + A,
\end{equation*}
where $A$ is an integration constant. It vanishes since $V^1$ and its derivative have to stay finite at $\eta = 0$. Using the boundary condition at $\eta = 1$, yields the expression for $G^1$:
\begin{equation}
\label{eq1G1}
G^1 = - U\dz{C}.
\end{equation}
We now use this to derive an expression for $V^1$, and integrate once more. The integration constant is found using the fact that the average over the cross section of $V^1$ is zero, $\int V^1 \eta \,d\eta = 0$, such that :
\begin{equation}
\label{eq1V1}
V^1 = \frac{a^2}{4\kappa} U\dz{C} \left[ \eta^2 - \frac{\eta^4}{2} - \frac{1}{3} \right].
\end{equation}
\paragraph{2nd order expansion}
Now we solve for $V^2$. The defining equations are:
\begin{equation*}
\begin{cases}
\frac{\kappa}{a^2 \eta} \de \left( \eta \dev{V^2} \right) = \dt{V^1} + \dC{V^0}G^2 + \dCz{V^1} \dz{G^1}+ \dC{V^1} G^1 + u\dz{V^1} - \kappa \dzz{V^0}\\
\dev{V^2}\big|_{\eta = 1} = 0 
\end{cases}
\end{equation*}
Now we use Eq.~\eqref{eq1V1} and the expression for $u$; and integrate once:
\begin{equation*}
\begin{split}
 \de \left( \eta \dev{V^2} \right) &= 0 + \frac{a^2}{\kappa}\bigg( G^2 \eta - U^2\dzz{C} \frac{a^2}{4\kappa}\eta \left[\eta^2 - \frac{\eta^4}{2} - \frac{1}{3} \right] + 0 \\
 &+ 2\frac{U^2a^2}{4\kappa} \dzz{C} \eta (1- \eta^2) \left[ \eta^2 - \frac{\eta^4}{2} - \frac{1}{3} \right] - \kappa \eta \dzz{C} \bigg)\\
 \eta \dev{V^2} &=  \frac{a^2}{\kappa}\bigg( G^2 \frac{\eta^2}{2} - U^2\dzz{C} \frac{a^2}{4\kappa} \left[\frac{\eta^4}{4} - \frac{\eta^6}{12} - \frac{\eta^2}{6} \right]\\
&+ \frac{U^2a^2}{2\kappa} \dzz{C} \left[ -\frac{\eta^2}{6} + \frac{\eta^4}{3}-\frac{\eta^6}{4}+\frac{\eta^8}{16} \right] - \kappa \frac{\eta^2}{2} \dzz{C} \bigg),
\end{split}
\end{equation*}
where the integration constant vanishes since $V^2$ has to be finite at $\eta = 0$, similarly to the previous integration. Using the boundary condition for $V^1$ yields:
\begin{equation}
\label{eq1G2}
G^2 = \kappa \left( 1 + \frac{a^2 U^2}{48 \kappa^2} \right) \dzz{C}.
\end{equation}
Combining this result with Eq.~\eqref{eq1G1} and~\eqref{eq1G2} we find the 2nd order approximation for the cross-sectionnally averaged concentration $C$:
\begin{equation}
\label{TD1}
 \dt{C} = -U\dz{C} +  \kappa \left( 1 + \frac{a^2 U^2}{48 \kappa^2} \right) \dzz{C}
\end{equation}
Eq.~\eqref{TD1} is exactly the Taylor dispersion equation. 
\subsection{Taylor Dispersion with solute absorption conditions at the wall}
\label{sec:Absorption}
\subsubsection*{Invariant manifold method with absorption in the steady uniform contracting channel}
We consider the problem in the steady uniform case again but this time with absorption at the boundaries.
The boundary condition at the boundaries of the channel with absorption writes:
\begin{equation*}
\partial_{\eta} c|_{\eta = 1} + \gamma' c|_{\eta = 1}= 0.
\end{equation*}
where $\gamma' = \gamma \frac{a}{\kappa}$ and $\gamma$ is the surface absorption coefficient in $\mathrm{m .\, s^{-1}}$.  In the following we drop the prime notation for $\gamma$ and keep it non-dimensional. 

\paragraph{0th order expansion}
As before we have $V^0 = C$ the mean concentration over the cross-section.

\paragraph{1st order expansion}
Now we solve for $V^1$. It verifies:
\begin{equation*}
\begin{cases}
\frac{\kappa}{a^2 \eta} \de \left( \eta \dev{V^1} \right) =\dt{V^0} + \dC{V^0} G^1 + u\dz{V^0} \\
\dev{V^1}\big|_{\eta = 1} = - \gamma V^{0}|_{\eta = 1} 
\end{cases}
\end{equation*}
We use the expression of $u(\eta)$, $V^0$:
\begin{equation*}
\frac{\kappa}{a^2 \eta} \de \left( \eta \dev{V^1} \right)  =  G^1  + 2 U \left(1 -\eta^2\right) \dz{C} 
\end{equation*}
and integrating once and using the boundary conditions we find $G^1$:
\begin{equation}
\label{eq1G1}
G^1 = - U \dz{C} - 2 C \gamma \frac{\kappa}{a^2}  
\end{equation}
where we notice that $G^1$ now also contains some sink term due to absorption at the boundaries. Reporting in the integral expression for $V^1$, and using the fact that the cross-sectional average of $V^1$ has to vanish yields the following expression:
\begin{equation}
V^1 = \frac{1}{4} (1 - 2 \eta^2) \gamma C +  \dz{C} \frac{U a^2}{24 \kappa} \left( -2 + 6 \eta^2 - 3\eta^4 \right).
\end{equation}

\paragraph{2nd order expansion}
Now we solve for $V^2$. It verifies:
\begin{equation*}
\begin{cases}
\frac{\kappa}{a^2 \eta} \de \left( \eta \dev{V^2} \right) = \dt{V^1} + \dC{V^0} G^2 + \dCz{V^1} \dz{G^1}+ \dC{V^1} G^1 + u\dz{V^1} - \kappa \dzz{V^0}\\
\dev{V^2}\big|_{\eta = 1} = - \gamma V^1|_{\eta = 1}
\end{cases}
\end{equation*}
Now we integrate one first time and use the boundary condition on $V^2$ gives:
\begin{equation}
G^2 = + \frac{1}{2} \gamma^2 \frac{\kappa}{a^2} C  - \frac{1}{6} \gamma U \dz{C}  + \kappa \left( 1 + \frac{a^2 U^2}{48 \kappa^2} \right) \dzz{C}.
\end{equation}
To compare the expansion with other works it is interesting to search for the next order in the expansion. We report just the order 3 term here:
\begin{equation}
G^3 = -\frac{1}{12} \gamma^3 \frac{\kappa}{a^2} C + \frac{1}{48} \gamma^2 U \dz{C} + \frac{a^2 U^2}{48 \kappa^2} \dzz{C} \frac{\gamma}{30} - \frac{a^2 U^2}{2880 \kappa^2} a^2 U C_{zzz}  
\end{equation}
And order 4 term:
\begin{equation}
G^4 = +\frac{1}{192} \gamma^3 \frac{\kappa}{a^2} C + \frac{1}{288} \gamma^2 U \dz{C} + \frac{a^2 U^2}{48 \kappa^2} \dzz{C} \frac{37\gamma^2}{360} + ... 
\end{equation}

\subsubsection*{Heuristic derivation for absorption at the boundaries}

From heuristic derivations of Taylor dispersion with absorption such as in~\cite{Meigel:2018hw} we have\footnote{Note that in~\cite{Meigel:2018hw} there is a typo in the formula given that we have updated here.}
\begin{equation}
\frac{\partial C}{\partial t} = - 2\frac{4 \gamma}{4 + \gamma}  \frac{\kappa}{a^2}  C - \frac{12 + 5\gamma}{12 + 3 \gamma} U \frac{\partial C}{\partial z} + \kappa \left(1 + \frac{a^2 U^2}{48 \kappa^2} \frac{12 + \gamma}{12 + 3\gamma} \right)  \frac{\partial^2 C}{\partial z^2} 
\end{equation}
This gives the coefficients plotted in the main text (gray) that are not as good to reproduce reality. This can be explained in the following way. As the heuristic expansion is a 2nd order expansion (and neglects all higher order terms),  one should stop at 2nd order terms in $\gamma$ and in $\frac{\partial .}{\partial z}$, which gives:
\begin{equation}
\frac{\partial C}{\partial t} = - 2 \gamma (1 - \gamma/4)  \frac{\kappa}{a^2}  C - (1 + \gamma/6) U \frac{\partial C}{\partial z} + \kappa \left(1 + \frac{a^2 U^2}{48 \kappa^2} \right)  \frac{\partial^2 C}{\partial z^2} 
\end{equation}
which is absolutely correct and corresponds to the invariant manifold expansion. 

\subsection{Taylor Dispersion in a straight channel with different flow profiles}

\subsubsection{Flow slip at the channel wall}
\label{sec:slip}
We consider the same problem as before, in the simplest setting where $\dt{a} = 0$ and $\dz{a} = 0$. 
This time, we allow for slippage at the boundaries of the channel. In other words, instead of having the standard Poiseuille profile, we impose:
\begin{equation*}
\dr{u}\bigg|_{r = a} = - \frac{u(r=a)}{b}
\end{equation*}
where $b > 0$ is a length called the {\it slip length}.
If we solve for the flow in the channel we find then:
\begin{equation*}
u = 2U_{0} \alpha \left( 1 - \frac{ \eta^2}{\alpha} \right)
\end{equation*}
with $\alpha = 1+2b/a$, such that the velocity at the interface is non zero. We also have that the cross-sectionally averaged flow is:
\begin{equation*}
U = (2\alpha -1) U_{0}
\end{equation*}

\paragraph{0th order expansion}
The assumptions on the flow profile do not change the other boundary conditions such that $V^0 = C$. 

\paragraph{1st order expansion}

Now we solve for $V^1$. It verifies:
\begin{equation*}
\begin{cases}
\frac{\kappa}{a^2 \eta} \de \left( \eta \dev{V^1} \right) = \dt{V^0} + \dC{V^0}G^1 + u\dz{V^0} \\
\dev{V^1}\big|_{\eta = 1} = 0 
\end{cases}
\end{equation*}
We use the expression of $u(\eta)$ and integrate once with respect to $\eta$:
\begin{equation*}
\eta \dev{V^1} = \frac{a^2}{\kappa} \left( \frac{\eta^2}{2} G^1 + 2U \left[ \frac{\eta^2}{2} - \alpha \frac{\eta^4}{4} \right] \dz{C} \right).
\end{equation*}
Note there that we did not add any integration constant. Automatically this constant is set to zero, otherwise, once divided by $\eta$, it would yield divergences in $\dev{V^1}$. 
We now use the boundary condition in $\eta = 1$, this yields the expression of $G^1$:
\begin{equation}
\label{eq1G1b}
G^1 = - U \dz{C} ( 2 - \frac{1}{\alpha} )
\end{equation}
We now use this to derive an expression for $V^1$. We find:
\begin{equation*}
\begin{split}
 \dev{V^1} &= \frac{a^2}{\kappa}U \alpha^{-1} \dz{C} \left( \frac{\eta}{2}  - \frac{\eta^2}{2} \right) \\
 V^1 &= B + \frac{a^2}{\kappa}  U  \alpha^{-1} \dz{C}  \left( \frac{\eta^2}{4}  - \frac{\eta^4}{8}\right) 
\end{split}
\end{equation*}
and using the fact that the average over the cross section of $V^1$ is zero, $\int V^1 \eta d\eta = 0$ we find $B = -\frac{a^2}{12 \kappa} U\alpha^{-1}  \dz{C}$, and thus:
\begin{equation}
\label{eq1V1b}
V^1 = \frac{a^2}{4\kappa} U \alpha^{-1} \dz{C} \left[ \eta^2 - \frac{\eta^4}{2} - \frac{1}{3} \right]
\end{equation}

\paragraph{Step 2}
Now we solve for $V^2$. It verifies:
\begin{equation*}
\begin{cases}
\frac{\kappa}{a^2 \eta} \de \left( \eta \dev{V^2} \right) = \dt{V^1} + \dC{V^0}G^2 + \dCz{V^1} \dz{G^1}+ \dC{V^1} \dzz{G^1} + u\dz{V^1} - \kappa \dzz{V^0}\\
\dev{V^2}\big|_{\eta = 1} = 0 
\end{cases}
\end{equation*}
Now we use Eq.~\eqref{eq1V1} and the expression for $u$; and integrate once:

Using the boundary condition on $V^2$ gives:
\begin{equation*}
\begin{split}
0 & = G^2 \frac{1}{2} - \alpha^{-1}(2 - \alpha^{-1}) U^2\dzz{C} \frac{a^2}{4\kappa} \left[\frac{1}{4} - \frac{1}{12} - \frac{1}{6} \right]\\
&+ \frac{U^2a^2}{2\kappa} \alpha^{-1} \dzz{C} \left[ (\frac{1}{4} - \frac{1}{12} - \frac{1}{6}) -  \alpha^{-1}  \left[ \frac{1}{6} - \frac{1}{16} - \frac{1}{12} \right] \right] - \kappa \frac{1}{2} \dzz{C} \\
0 & = G^2 \frac{1}{2} - 0 \\
&- \frac{U^2a^2}{2\kappa} \alpha^{-1} \dzz{C}  \alpha^{-1}  \left[ \frac{8}{6} - \frac{3}{48} - \frac{4}{48} \right] - \kappa \frac{1}{2} \dzz{C} 
 \\
 G^2 &= \frac{U^2a^2}{48\kappa} \alpha^{-1} \dzz{C}  \alpha^{-1} + \kappa \dzz{C} 
\end{split}
\end{equation*}

\begin{equation}
\label{eq1G2b}
G^2 = \kappa \left( 1 + \frac{a^2 U_0^2}{48 \kappa^2} \right) \dzz{C}
\end{equation}
and with no further ado, assembling Eq.~\eqref{eq1G1b} and~\eqref{eq1G2b} yields the following second order approximate equation for the cross-sectionally averaged concentration $C$:
\begin{equation}
\label{TD1bappendix}
\dt{C} = -U_0(2\alpha - 1)\dz{C} +  \kappa \left( 1 + \frac{a^2 U_0^2}{48 \kappa^2} \right) \dzz{C}
\end{equation}
Eq.~\eqref{TD1bappendix} is exactly Eq.~\eqref{TD1b} of the main text. 
\subsubsection{Arbitrary flow profile}
\label{sec:arbitrary}
We consider the same problem as before, in the simplest setting where $\dt{a} = 0$ and $\dz{a} = 0$. 
The flow profile is some function of the coordinate $\eta$, $u(\eta)$. We do not impose any condition on it for now (except standard regularity). The orthogonal flow is still zero. 

\paragraph{0th order expansion} This flow profile does not change the other boundary conditions such that $V^0 = C$. 

\paragraph{1st order expansion} 
Now we solve for $V^1$. It verifies:
\begin{equation*}
\begin{cases}
\frac{\kappa}{a^2 \eta} \de \left( \eta \dev{V^1} \right) = \dt{V^0} + \dC{V^0}G^1 + u\dz{V^0} \\
\dev{V^1}\big|_{\eta = 1} = 0 
\end{cases}
\end{equation*}
We use the expression of $u(\eta)$ and integrate once with respect to $\eta$:
\begin{equation*}
\eta \dev{V^1} = \frac{a^2}{\kappa} \left( \frac{\eta^2}{2} G^1 +  \left[ \int_0^\eta u(x) x dx \right] \dz{C} \right)
\end{equation*}
and using the boundary condition in $\eta = 1$, this yields the expression of $G^1$:
\begin{equation}
\label{eq1G1c}
G^1 = - U \dz{C} 
\end{equation}
where $U = 2 \int_0^1 u(x) x dx$ is the mean flow. 
We now use this to derive an expression for $V^1$. We find:
\begin{equation*}
\begin{split}
 \dev{V^1} &= \frac{a^2}{\kappa}\dz{C} \left( -\frac{\eta}{2} U  + \eta^{-1} \int_0^{\eta} u(x) x dx \right) \\
 V^1 &= B +\frac{a^2}{\kappa}\dz{C} \left( -\frac{\eta^2}{4} U  + \int_0^{\eta} \frac{1}{y} \int_0^{y} u(x) x dx \right).
\end{split}
\end{equation*}
At this stage let's introduce $\tilde{u}(\eta) =  \int_0^{\eta} \frac{1}{y} \int_0^{y} u(x) x dx$. We use the fact that the average over the cross section of $V^1$ is zero, $\int V^1 \eta d\eta = 0$ we find $B = \frac{a^2}{8 \kappa} U \dz{C} -  \frac{a^2}{ \kappa} \overline{\tilde{u}} \dz{C}$, and thus:
\begin{equation}
\label{eq1V1c}
V^1 = \frac{a^2}{4\kappa} U  \dz{C} \left[ \frac{1}{2} - \eta^2 \right] +  \frac{a^2}{\kappa}  \dz{C} \left[ \tilde{u}  - \overline{\tilde{u}} \, \right]
\end{equation}

\paragraph{2nd order expansion}
Now we solve for $V^2$. It verifies:
\begin{equation*}
\begin{cases}
\frac{\kappa}{a^2 \eta} \de \left( \eta \dev{V^2} \right) = \dt{V^1} + \dC{V^0}G^2 + \dCz{V^1} \dz{G^1}+ \dC{V^1} \dzz{G^1} + u\dz{V^1} - \kappa \dzz{V^0}\\
\dev{V^2}\big|_{\eta = 1} = 0 
\end{cases}
\end{equation*}
Now we use Eq.~\eqref{eq1V1c} and integrate once:
\begin{equation*}
\begin{split}
 \eta \dev{V^2} &=  \frac{a^2}{\kappa}\bigg( G^2 \frac{\eta^2}{2} - U^2\dzz{C} \frac{a^2}{\kappa} \int_0^{\eta} d\eta \left[\frac{\eta}{8} - \frac{\eta^3}{4} + \frac{\tilde{u}(\eta)}{U} \eta - \frac{\overline{\tilde{u}}(\eta)}{U} \eta \right] \\
&+ \frac{Ua^2}{\kappa}\dzz{C} \int_0^{\eta} \left[ \frac{u(\eta)\eta}{8} - \frac{u(\eta)\eta^3}{4} + \frac{u(\eta)\tilde{u}(\eta)\eta}{U} - u(\eta)\eta\frac{\overline{\tilde{u}}}{U}  \right] - \kappa \frac{\eta^2}{2} \dzz{C} \bigg)
\end{split}
\end{equation*}

Using the boundary condition on $V^2$ after some straightforward calculus:
\begin{equation*}
\begin{split}
 G^2 &= \frac{U^2a^2}{48\kappa} \dzz{C} I[u]  + \kappa \dzz{C} 
\end{split}
\end{equation*}
where
\begin{equation}
\label{IdeU}
I[u] = 48 \int_0^1 2\eta d\eta \left[ \frac{u(\eta)}{U} - 1 \right] \left[ \frac{\tilde{u}(\eta)}{U} - \frac{\eta^2}{4} \right]
\end{equation}
such that we get the following 2nd order approximate equation for the cross sectionnally averaged concentration $C$:
\begin{equation}
\label{TD1c}
 \dt{C} = -U\dz{C} +  \kappa \left( 1 + \frac{a^2 U^2}{48 \kappa^2} I[u] \right) \dzz{C}
\end{equation}
Eq.~\eqref{TD1c} is exactly Eq.~\eqref{eq:arbitrary} in the main text. 
\subsubsection{Optimal flow profile}
\label{sec:optimal}
First, we make the change of variable: $\psi(\eta) = \frac{u(\eta)}{U} - 1$. $\overline{\psi} = 0$ and measures the deviation of the velocity profile $u$ to a plug flow. One can rewrite Eq.~\eqref{IdeU} as:
\begin{equation}
I[u] = 96 \int_0^1 \frac{1}{x} \left[ \int_0^x x' \psi(x') dx' \right]^2 =  96 \int_0^1 \frac{1}{x} \int_0^x x' \psi(x') dx' \int_0^x x'' \psi(x'') dx'' .
\end{equation}
There the space of integration is for all $x \geq (x',x'')$ such that we can also rewrite:
\begin{equation}
\begin{split}
I[u] &=  96 \int_0^1 dx' \int_0^1 dx'' \int_{\max(x',x'')}^1\frac{ dx }{x} x' \psi(x')  x'' \psi(x'') \\
&= - \iint_0^1 dx' dx''  \psi(x')   \psi(x'')  x' x'' \ln \left( \max(x',x'')\right)
\end{split}
\end{equation}

We search for the optimal flow profile $v$, that maximizes $I[u]$. We will impose a few conditions. 
First, the average over the cross-section has to vanish: 
\begin{equation}
\int_0^1 \psi(x) x dx = 0
\label{condMean}
\end{equation} 
and second, we impose that the $L^2$ norm is bounded, namely that the deviation to the plug flow has a finite strength:
\begin{equation}
\int_0^1 \psi(x)^2 x dx = \alpha
\label{condAlpha}
\end{equation}
where $\alpha = 1/6$ is taken from the Poiseuille flow reference. 

The problem then amounts to finding an extremal solution of the following lagrangian:
\begin{equation}
\mathcal{L}[v] = \iint dx dy \psi(x) \psi(y) xy \ln (\max (x,y)) + \lambda_0 \int \psi(x) x dx + \lambda_1 \left( \int \psi(x)^2 x dx - \alpha \right)
\end{equation}
To do so we search for $v$ such that for any function $h(x)$ we have $\mathcal{L}[\psi+h] - \mathcal{L}[\psi] = 0$. This condition writes:
\begin{equation}
\int_0^1 h(x) dx \left[ \int 2 dy xy \ln(\max(x,y)) \psi(y) + \lambda_0 x + 2\lambda_1 x \psi(x) \right] = 0 , \, \forall h(x).
\end{equation}
This is equivalent to searching $\psi$ such that $\forall x$:
\begin{equation}
\int 2 dy xy \ln(\max(x,y)) \psi(y) + \lambda_0 x + 2\lambda_1 x \psi(x) = 0
\end{equation}
and as this equation is true $\forall x$ it is true in particular for $x\neq 0$:
\begin{equation}
\begin{split}
&\int 2 dy y \ln(\max(x,y)) \psi(y) + \lambda_0 + 2\lambda_1 \psi(x) = 0 \\
\Leftrightarrow & \int_0^x 2 dy y \psi(y) \ln (x) + \int_x^1 2 dy y \psi(y) \ln (y) + \lambda_0 + 2\lambda_1 \psi(x) = 0 
\end{split}
\end{equation}
and since $\int_0^1 \psi(y) y dy = 0$ we find:
\begin{equation}
\int_x^1 2 dy y \psi(y) ( \ln (y) - \ln(x)) + \lambda_0 + 2\lambda_1 \psi(x) = 0, \, \forall x
\end{equation}
As this is true $\forall x$, it is true for the derivative of this function as well, 
\begin{equation}
-\frac{1}{x}\int_x^1 dy y \psi(y) + \lambda_1 \psi'(x) = 0, \, \forall x
\end{equation}
multiplying by $x$ and performing the derivative one more time gives
\begin{equation}
x \psi(x) + \lambda_1 \psi''(x)x + \lambda_1 \psi'(x) = 0, \, \forall x
\end{equation}
which is a second order partial differential equation. Assuming that $\psi(x)$ is sufficiently regular to be expandable as series:
$\psi(x) = \sum_n a_n x^n$, one easily finds that the optimal solution $\psi$ is a Bessel function:
\begin{equation}
\psi(x) = \sum_{p=0}^{\infty} \left( \frac{x}{\sqrt{\lambda_1}}\right)^{2p} (-1)^p \frac{1}{2^{2p}(p!)^2} a_0 = a_0 J_0\left( \frac{x}{\sqrt{\lambda_1}} \right).
\end{equation}
To close the problem one has to find $\lambda_1$ such that Eq.~\eqref{condMean} is verified and $a_0$ such that Eq.~\eqref{condAlpha} is verified. $\lambda_1$ has multiple solutions:
\begin{equation}
\begin{cases}
\int_0^1  a_0 J_0\left( \frac{x}{\sqrt{\lambda_1}} \right) x dx = 0 \\
\int_0^1 a_0^2 J_0^2\left( \frac{x}{\sqrt{\lambda_1}} \right) x dx = \alpha
\end{cases}
\end{equation} The first equality has multiple roots in $\lambda_1$. The largest root is $\lambda_1 \simeq 0.068$ and actually corresponds to the optimal value, while the other roots corresponds to smaller values of $I[u]$. This value of $\lambda_1$ gives $a_0 \simeq 1.43$ in the second equality. 
\subsection{Taylor Dispersion in a non uniform time contracting channel}
We focus now on the case where the radius of the channel $a(z,t)$ depends on the axial position along the channel $z$ and time $t$. Before we can start solving the problem in this case, some ideas must be set up. We highlight at this point that the non dimensional variable $\eta(z,t) = \frac{r}{a(z,t)}$ depends on $z$ and time $t$. 

\paragraph{Problem set up}

We consider that the flow is quasi-Poiseuille, such that we have a similar expression for the longitudinal velocity $u$, but there is an axial component $v$ that can be computed easily by solving the zero divergence. 
\begin{equation*}
\begin{cases}
u = 2U (1 - \eta^2) \\
v = 2 \dz{a} U (\eta - \eta^3) +\dt{a} \eta (2 - \eta^2)
\end{cases}
\end{equation*} 
Moreover, one must not forget that considering that the flow is incompressible yields the following useful relationship between the mean flow and the cross section area:
\begin{equation*}
\partial_z(a^2 U) = - 2 a \dt{a} \Leftrightarrow 2a\dz{a} U + a^2 \dz{U} = -2a \dt{a}
\end{equation*}

The no solute flux boundary condition at the channel boundary still writes (at first order in $\dz{a}$):
\begin{equation*}
\frac{\kappa}{a} \frac{\partial C}{\partial \eta} - \kappa \frac{\partial a}{\partial z} \frac{\partial C}{\partial z} = 0
\end{equation*}
Note that in this expression, $\dt{a}$ does not appear because the condition no solute flux boundary condition may be written in the moving reference frame of the boundary. 

\paragraph{0th and 1st order expansion}

The 0th and 1st order of the expansion do not change as compared to the straight steady case and thus we may start off with: $V^0 = C$, $G^1 = - U \dz{C}$ and $V^1 = \frac{a^2}{4\kappa} U\dz{C} \left[ \eta^2 - \frac{\eta^4}{2} - \frac{1}{3} \right]$.

\paragraph{2nd order expansion}
Now we solve for $V^2$. It verifies:
\begin{equation*}
\begin{cases}
\frac{\kappa}{a^2 \eta} \de \left( \eta \dev{V^2} \right) = \dt{V^1} + \dC{V^0} G^2 + \dCz{V^1} \dz{G^1}+ \dC{V^1} \dzz{G^1}+ u\dz{V^1}+ \frac{v}{a} \dev{V^1} - \kappa \dzz{V^0}\\
\frac{1}{a}\dev{V^2}\big|_{\eta = 1} = \dz{a} \dz{C}
\end{cases}
\end{equation*}
As compared to the previous case, we have to pay special attention to derivatives in space and time especially because the non-dimensional coordinate $\eta = r/a(z,t)$ is dependent in space and time as well:
\begin{equation*}
\begin{split}
\dz{V_1} &= \frac{\dzz{C}}{4\kappa} Ua^2 \left[ \eta^2 - \frac{\eta^4}{2} - \frac{1}{3} \right] + \frac{\dz{C}}{4\kappa} \eta^2 a^2 \dz{U} \\
 &- \frac{r^4 \dz{C}}{8\kappa} \left( \frac{\dz{U}}{a^2} - \frac{2U \dz{a}}{a^3} \right) - \frac{\dz{C}}{12\kappa} (-2a\dt{a}) \\
 \dz{V_1} &=  \frac{\dzz{C}}{4\kappa} Ua^2 \left[ \eta^2 - \frac{\eta^4}{2} - \frac{1}{3} \right] + \frac{\dz{C}}{4\kappa} \eta^2 ( - 2 a \dz{a} U - 2 a \dt{a}) \\
 & - \frac{\eta^4 \dz{C}}{8\kappa} \left( - 4 a \dz{a} U - 2 a \dt{a} \right) - \frac{\dz{C}}{12\kappa} (-2a \dt{a}) \\
\dz{V_1} &=  \frac{\dzz{C}}{4\kappa} Ua^2 \left[ \eta^2 - \frac{\eta^4}{2} - \frac{1}{3} \right] - \frac{\dz{C}}{2\kappa}a \dz{a} U \left(\eta ^2 -  \eta^4 \right) -  \frac{\dz{C}}{2\kappa}a \dt{a} \left[\eta^2 - \frac{\eta^4}{2} - \frac{1}{3} \right]
\end{split}
\end{equation*}
And also to $\dt{V^1}$ that is non zero here:
\begin{equation*}
\begin{split}
\dt{V^1} &= \frac{\partial}{\partial t} \left( \frac{\dz{C}}{4\kappa} \left( Ur^2 - \frac{r^4U}{2a^2} - \frac{1}{3}a^2U \right) \right)\\
\dt{V^1} &= \frac{\dz{C}}{4\kappa}a^2 \dt{U} \left( \eta^2 - \frac{\eta^4}{2} - \frac{1}{3} \right) + \frac{\dz{C}}{4\kappa}U a \dt{a} \left( \eta^4 - \frac{2}{3} \right)
\end{split}
\end{equation*}
We now assemble everything and integrate once to find $G^2$ at the boundary. For simplicity, in the following derivation we include by a symbolic $\tilde{f}(\eta) =  \left[\eta^2 - \frac{\eta^4}{2} - \frac{1}{3} \right] $. All terms that are not multiplied by another function of $\eta$ are forgotten at the second line because they cancel out in the derivation:
\begin{equation*} 
\begin{split}
\frac{\kappa}{a^2 \eta} \de \left( \eta \dev{V^2} \right) &= \frac{\dz{C}}{4\kappa}a^2 \dt{U} \tilde{f}(\eta) + \frac{\dz{C}}{4\kappa}U a \dt{a} \left( \eta^4 - \frac{2}{3} \right) + G^2  - U^2\dzz{C} \frac{a^2}{4\kappa} \tilde{f}(\eta) + 0 \\
 & - \kappa \dzz{C} - \frac{a^2}{4\kappa} \dz{U} U \dz{C} \tilde{f}(\eta)  \\
  &+ 2U(1-\eta^2) \left( \frac{a^2}{4\kappa} U \dzz{C} \tilde{f}(\eta) - \frac{a \dz{a} U \dz{C}}{2\kappa}  (\eta^2 - \eta^4)  -  \frac{\dz{C}}{2\kappa}a \dt{a} \tilde{f}(\eta) \right) \\
  &+ \left( \frac{ 2 \dz{a} U (\eta - \eta^3)}{a} + \frac{\dt{a} \eta(2-\eta^2)}{a}\right)\frac{a^2}{2\kappa} U\dz{C} \eta (1 - \eta^2) 
\end{split}
\end{equation*}
\begin{equation*} 
\begin{split}
  \frac{\kappa}{a^2 \eta} \de \left( \eta \dev{V^2} \right) &= G^2  - \kappa \dzz{C} + \dzz{C} \frac{a^2U^2 }{2\kappa} (1-\eta^2) \left[ \eta^2 - \frac{\eta^4}{2} - \frac{1}{3} \right]  \\
  &- \frac{a \dz{a}}{\kappa}U^2 \dz{C} \left(  (1 - \eta^2)(\eta^2 - \eta^4) - (\eta - \eta^3)\eta(1-\eta^2)\right) \\
&+ \frac{\dz{C} }{2\kappa}a \dt{a} U^2 \left( \frac{1}{2} \left(\eta^4 - \frac{2}{3} \right) + (1-\eta^2) \left( -2 [\eta^2 - \frac{\eta^4}{2} - \frac{1}{3} ] + \eta^2 (2 - \eta^2) \right) \right)  \\
 \de \left( \eta \dev{V^2} \right) &=\frac{a^2}{\kappa} G^2 \eta  - a^2\eta \dzz{C} + \dzz{C} \frac{a^4U^2 }{2\kappa^2} \eta (1-\eta^2) \left[ \eta^2 - \frac{\eta^4}{2} - \frac{1}{3} \right]  \\
&  + \frac{\dz{C} }{2\kappa^2}a^3 \dt{a} U \left( \frac{1}{3} - \frac{\eta^4}{2} - \frac{2\eta^2}{3}\right) \\
  \eta \dev{V^2} &=\frac{a^2}{2\kappa} G^2 \eta^2 - \frac{a^2}{2} \eta^2 \dzz{C} + \dzz{C} \frac{a^4U^2 }{2\kappa^2}\left[ -\frac{\eta^2}{6} + \frac{\eta^4}{3}-\frac{\eta^6}{4}+\frac{\eta^8}{16} \right] \\
  &+ \frac{\dz{C}}{2\kappa^2}a^3\dt{a} U \left( \frac{\eta^2}{6} + \frac{\eta^6}{12} - \frac{\eta^4}{6} \right)
\end{split}
\end{equation*}

Now we use the boundary condition:
\begin{equation*}
a \dz{a} \dz{C} =\frac{a^2}{2\kappa} G^2    - \frac{a^2}{2}  \dzz{C} - \dzz{C} \frac{a^4U^2 }{96\kappa^2} + \frac{\dz{C} a^3 \dt{a} U^2}{24 \kappa^2}
\end{equation*}
and thus:
\begin{equation}
G^2  = \frac{2 \kappa \dz{a}}{a} \dz{C} + \kappa \dzz{C} + \frac{a^2U^2}{48\kappa} \dzz{C}   - \frac{\dz{C} a \dt{a} U^2}{12 \kappa}
\end{equation}
And thus if we assemble the parts to get the partial differential equation for $C$ we find:
\begin{equation*}
 \dt{C} = -U\dz{C} + \frac{2 \kappa \dz{a}}{a} \dz{C} + \kappa \left( 1 + \frac{a^2 U^2}{48 \kappa^2} \right) \dzz{C} - \frac{a  \dt{a} U}{24 \kappa} \dz{C}
\end{equation*}
Note at this point that $\dz{a^2U} = -2 \dt{a} a$, and thus one can rewrite the previous equation in the simpler form : 
\begin{equation}
 \dt{a^2C} = \frac{\partial}{\partial z} \left( - a^2UC +  a^2 \kappa \left( 1 + \frac{a^2 U^2}{48 \kappa^2} \right) \dz{C} \right)
\end{equation}
\subsection{Effect of pumping on dispersion}
\label{sec:longtime}
We start with the differential equation on $\rho$ 
\begin{equation}
\begin{split}
& \dt{\rho}  = \frac{\partial}{\partial \xi} \left(-\tilde{U}(\xi)\rho + \tilde{\kappa}(\xi)  \frac{\partial\rho}{\partial \xi}\right),\\
\mathrm{with}  \,\, \tilde{U}(\xi) =  \frac{\omega}{k} - U + &\frac{\kappa}{a^2}\frac{\partial a^2}{\partial \xi} \left( 1 + \frac{a^2 U^2}{48 \kappa^2} \right) \,\,  \mathrm{and} \, \, \tilde{\kappa}(\xi) = \kappa \left( 1 + \frac{a^2 U^2}{48 \kappa^2} \right)
\end{split}
\end{equation}
At second order in $\phi$ we have:
\begin{equation}
\tilde{U}(\xi) =  v_{p} - \phi v_{p} \cos(k \xi) ( 1 - \phi \cos(k \xi)) - \kappa k \phi \sin (k \xi) ( 1 - \phi \cos(k\xi)) + O(\phi^3) 
\end{equation}
with the phase velocity $v_{p} = \omega/k$ and
\begin{equation}
\tilde{\kappa}(\xi) = \kappa + \frac{a^2 v_p^2}{48 \kappa}\phi^2 \cos^2 (k\xi) +  O(\phi^3) 
\end{equation}
Clearly the problem is periodic with $\tilde{U}$ and $\tilde{\kappa}$ periodic in $L = 2\pi/k$. We are facing a one-dimensional advection-diffusion problem, for which the long time diffusion coefficient writes
\begin{equation}
\kappa_{\rm eff} = \lim_{t\rightarrow \infty} \frac{\langle x^2(t) \rangle - \langle x(t) \rangle^2}{2t}
\end{equation}
where $x(t)$ represents a stochastic variable, typically the position of the particle. The long time diffusion coefficient can be calculated either by the method of moments of first passage times~\cite{reimann2001giant} or by investigating probability distributions as in~\cite{guerin2015kubo}.  It is simply expressed as
\begin{equation}
\kappa_{\rm eff} =  (k^2 L^2) \frac{\int_0^L dx \, \kappa(x) (I_+(x)^2 I_-(x) + I_-(x)^2I_+(x))}{(\int_0^L dx I_+(x))^3+(\int_0^L dx I_-(x))^3}
\end{equation}
where 
\begin{equation}
I_{+}(y) = \exp( \Gamma(y)) \int_y^{+\infty} dx \frac{\exp(-\Gamma(x))}{\tilde{\kappa}(x)}
\end{equation}
\begin{equation}
I_{-}(y) = \exp( -\Gamma(y)) \int_{-\infty}^y  dx \frac{\exp(\Gamma(x))}{\tilde{\kappa}(x)}
\end{equation}
and
\begin{equation}
\begin{split}
\Gamma(y) = \int_0^y \frac{\tilde{U}(y')}{\tilde{\kappa}(y')} dy' .
 \end{split}
\end{equation}
Since we work in the limit $\phi \ll 1$ (small deformations) it is possible to expand the exponentials present in $I_+$ and $I_-$. This allows to compute the integrals analytically and after some lengthy though easy computations, one finds:
\begin{equation}
\kappa_{\rm eff} = \kappa\left( 1 + \frac{\phi^2}{96} \frac{a_0^2 v_p^2}{\kappa^2} + \frac{\phi^2}{2} \left( \frac{3 v_p^2 - \kappa^2 k^2}{v_p^2 + \kappa^2 k^2}\right) \right).
\end{equation}
Exactly the result of the main paper. 

\sop{The effective drift may also be obtained
\begin{equation}
v_{\rm eff} = \lim_{t\rightarrow\infty} \frac{\langle x(t) \rangle}{t} = \frac{L}{\int_{0}^L I_+(y)}
\end{equation}
In this setting one finds at second order 
\begin{equation}
v_{\rm eff} = \frac{\omega}{k} + O(\phi^3).
\end{equation}
Note that this effective speed is taken in the moving referential, so one has to subtract the relative referential speed to get the drift in the original referential. Also, to get the appropriate effective drift in the original referential, one should add the mean input current from the boundary, which at order 2 writes $\frac{\phi^2}{2}\frac{\omega}{k}$ and therefore in the original referential one has
\begin{equation}
v_{\rm eff} = \frac{\phi^2}{2}\frac{\omega}{k} + O(\phi^3).
\end{equation}
} 

%

\end{document}